\documentclass[a4paper,cleveref,english, autoref, thm-restate]{lipics-v2021}

\pdfoutput=1 

\RequirePackage{ifthen}

\newcommand{\typeof}{1} %

\newcommand{\longversion}[1]{\ifthenelse{\equal{\typeof}{0}}{}{#1}}
\newcommand{\shortversion}[1]{\ifthenelse{\equal{\typeof}{0}}{#1}{}}
\newcommand{\longshortversion}[2]{\ifthenelse{\equal{\typeof}{0}}{#2}{#1}}

\usepackage{soul}
\usepackage{adjustbox}
\usepackage{tikz}
\usepackage{boxedminipage}
\usepackage{bussproofs} 
\usepackage{stmaryrd}
\usetikzlibrary{cd}
\usepackage{proof}

\newcommand{\und}[1]{\underline{#1}}
\newcommand{\la}{\lambda}
\newcommand{\app}{~}
\newcommand{\fst}{\pi_1}
\newcommand{\snd}{\pi_2}
\newcommand{\pair}[1]{\langle #1 \rangle}
\newcommand{\ifthen}[3]{\mathsf{if}~#1~\mathsf{then}~#2~\mathsf{else}~#3}
\newcommand{\true}{\mathsf{tt}}
\newcommand{\false}{\mathsf{ff}}
\newcommand{\sub}[2]{[#2/#1]}
\newcommand{\seq}[1]{\tilde{#1}}
\newcommand{\LanguageSet}[2]{\Lambda_{#1,#2}}
\newcommand{\LaCF}{\LanguageSet{\mathcal{C}}{\mathcal{F}}}
\newcommand{\type}[1]{\mathtt{#1}}
\newcommand{\B}{\type{B}}
\newcommand{\sem}[1]{\llbracket #1 \rrbracket}
\newcommand{\catname}[1]{\mathbf{#1}}
\newcommand{\rr}{\boldsymbol{r}}
\newcommand{\xx}{\boldsymbol{x}}

\newcommand{\set}[1]{\ \{ #1 \} }

\newcommand{\Comp}{\star}
\newcommand{\boxx}[1]{[#1]}
\newcommand{\LOHOL}{LOHOL}
\newcommand{\complem}[1]{#1^{\complement}}
\newcommand{\comprehfull}[2]{\set{#1 \mid #2}}
\newcommand{\compreh}[1]{\comprehfull{\Theta}{#1}}
\newcommand{\open}{\mathcal{O}}
\newcommand{\cont}{\mathcal{C}}
\newcommand{\p}{\vdash}
\newcommand{\pT}{\vdash^{\Theta}}
\newcommand{\pTT}{\vdash^{\Theta_1;\Theta_2}}
\newcommand{\AXC}[1]{\AxiomC{#1}}
\newcommand{\UIC}[1]{\UnaryInfC{#1}}
\newcommand{\BIC}[1]{\BinaryInfC{#1}}
\newcommand{\TIC}[1]{\TrinaryInfC{#1}}
\newcommand{\DP}{\DisplayProof}
\newcommand{\LRR}[2]{~\mathcal{R}^{#1}_{#2}~}
\newcommand{\LRRTh}[1]{\LRR{\Theta}{#1}}

\newcommand{\vvskip}{\medskip}
\newcommand\midd{\; \mbox{\Large{$\mid$}}\;}
\newenvironment{framed}[0]{\begin{boxedminipage}{\linewidth}}{\end{boxedminipage}}

\usepackage{nameref}

\makeatletter
\let\orgdescriptionlabel\descriptionlabel
\renewcommand*{\descriptionlabel}[1]{%
	\let\orglabel\label
	\let\label\@gobble
	\phantomsection
	\edef\@currentlabel{#1}%
	\let\label\orglabel
	\orgdescriptionlabel{#1}%
}
\makeatother

\hideLIPIcs  

\shortversion{
	\title{Open Higher-Order Logic}
}
\longversion{
	\title{Open Higher-Order Logic (Long Version)
	}
}

\author{Ugo Dal Lago}{Dipartimento di Informatica-Scienza e Ingegneria, Universit\`a di Bologna}{ugo.dallago@unibo.it}{https://orcid.org/0000-0001-9200-070X}{}
\author{Francesco Gavazzo}{Dipartimento di Informatica, Universit\`a di Pisa}{francesco.gavazzo@unipi. it}{https://orcid.org/0000-0002-2159-0615}{}
\author{Alexis Ghyselen}{Dipartimento di Informatica-Scienza e Ingegneria, Universit\`a di Bologna}{alexis.ghyselen@unibo.it}{https://orcid.org/0000-0001-9767-2011}{}

\authorrunning{U.~Dal Lago, F.~Gavazzo, A.~Ghyselen}
\Copyright{U.~Dal Lago, F.~Gavazzo, A.~Ghyselen}



\funding{This work was funded by the ERC CoG ''DIAPASoN'' GA 818616.}

\nolinenumbers 

\EventEditors{Bartek Klin and Elaine Pimentel}
\EventNoEds{2}
\EventLongTitle{31st EACSL Annual Conference on Computer Science Logic (CSL 2023)}
\EventShortTitle{CSL 2023}
\EventAcronym{CSL}
\EventYear{2023}
\EventDate{February 13--16, 2023}
\EventLocation{Warsaw, Poland}
\EventLogo{}
\SeriesVolume{252}
\ArticleNo{23}

\begin{document}

\maketitle

\begin{abstract}
	We introduce a variation on Barthe et al.'s higher-order logic in which formulas 
	are interpreted as predicates over \emph{open} 
	rather than \emph{closed} objects. This way, concepts which have an 
	intrinsically functional nature, like continuity, differentiability, or 
	monotonicity, can be expressed and reasoned about in a very natural way, 
	following the structure of the underlying program. We 
	give open higher-order logic in distinct flavors, and in particular in its  
	\emph{relational} and \emph{local} versions, the latter being tailored for 
	situations in which properties hold only in \emph{part} of the underlying 
	function's domain of definition.
\end{abstract}

\section{Introduction}

Reasoning about functional programs poses a whole series  of  
challenges due, in particular, to the presence of higher-order 
constructions. A class of  methodologies  particularly suitable for
compositional reasoning on such programs is that of 
\emph{type systems}, which can  be seen as lightweight formal 
methods  for the verification of  relatively simple properties, mainly 
safety ones. In the last  thirty years, it has become 
apparent that   properties beyond  mere safety are actually amenable 
to be verified through  types, e.g,  termination \cite{Coppo1978:IntersectionTypes,RowanPfenning2000:Intersection}, complexity bounds \cite{Hofmann2003:NonSizeIncreasing,DalLagoGaboardi2011:LinearDependentTypes}, 
and noninterference \cite{Pottier2002:NonInterference}. In all these cases, types serve as expressions 
meant to abstractly describe the input and output interfaces 
to functions. Various levels of abstractions in turn give rise to  
distinct levels of expressiveness, and  to type inference problems of 
varying degrees of difficulty. 

A somehow different, although related, approach is the one of program 
logics,  in which types are replaced or complemented by formulas 
written in a logical  language. This approach, pioneered by Floyd and 
Hoare in the context of  first-order imperative languages \cite{Floyd1993:HoareLogicBook,Hoare1969:HoareLogic}, is nowadays common also in the realm of higher-order programming 
languages \cite{Brady2013:Idris}, where it stands out for its expressive power. 
Indeed, relative completeness results \cite{Cook1978:RelativeCompleteness}, 
which  hold in many contexts within the realm program logics, are 
rarer in type systems.
 
A simple, yet powerful, form of program logic among those 
capable of dealing with  higher-order programs is \emph{higher-order 
logic}, as formulated by  Aguirre et al. in  a series of recent 
works \cite{Aguirre2017:RelationalLogic,Aguirre2020:RelationalLogics,AguirreetAl2018:RHOLMarkovChains,Aguirreetal2019:VerifHOProbPrograms}, most of which focusing on relational reasoning about 
higher-order programs. One common trait between the many  introduced 
dialects of higher-order logic is the fact that object 
programs are taken to be terms of a simply-typed  $\lambda$-calculus, 
while properties are  expressed in predicate logic. In the  words of 
Aguirre et al.~\cite{Aguirre2017:RelationalLogic}, higher-order logic 
\emph{can be understood as an attempt to internalize
the versatility of relational logical relations in a syntactic framework}.
Indeed, rules  of higher-order logic are built in such a way 
that syntax-directed reasoning about programs can be done 
following the rules provided by logical relations \cite{Plotkin1973:LogicalRelations,Statman1985:LogicalRelations}: any basic property 
at ground types is generalized at higher types by stipulating, 
conceptually, that related  functions should map related inputs  
to related outputs.  

As observed in several works \cite{Barthe2020:OpenLogicalRelations,JungTiuryn1993:OpenLambdaDefinability,StatonetAl2016:HigherOrder}, 
however, ordinary logical relations  
cannot easily  deal with properties which are higher-order extensions 
of \emph{first-order},  rather than ground, properties. This includes 
continuity, differentiability, or monotonicity properties. Indeed, 
what would be the base case in the (recursive) definition of the 
corresponding logical relation? Properties like continuity hold for 
functions and are meaningless if formulated for, say, real numbers. 
As we  show in Section~\ref{s:Reasoning} below, expectedly, similar
difficulties show  up when dealing with the same kind of properties 
in higher-order logic. Reasoning is indeed possible, but becomes 
cumbersome and difficult to be  carried out compositionally, i.e. in 
a syntax-directed way.  Going back to logical relations, however, 
there is a way out, which consists in switching to an \emph{open} 
version of logical relations in which  the base case is indeed the 
one of first-order types. This way, one is allowed  to start from, 
say, continuity and generalize it to higher-order types naturally. 

In this paper, we show that open logical relations can themselves be 
given a  formal counterpart in the realm of higher-order logic. We do 
so by introducing a variation on Aguirre et al.'s higher-order logic, 
called \emph{open higher-order logic}. The salient feature of this new system, 
compared to those from the literature, is the fact that the 
mathematical objects that proofs implicitly deal with, namely higher-order 
functions, are assumed to in turn depend on a sequence of ground global parameter $\Theta = \xx_1 : \B_1, \dots, \xx_n : \B_n$, hence the attribute ``open''. This way, a predicate $P$ about objects of type $\tau$ 
is taken as a subset of $\sem{\tau}^{\sem{\Theta}}$, rather than just $\sem{\tau}$.

Technically, the contributions of this paper are threefold:
\begin{itemize}
\item
	We define four concrete logical systems, all built around the 
	aforementioned ideas, and capable of dealing with formulas, programs, relations between programs (see Section~\ref{s:OHOL} for those three systems), and local reasoning (Section~\ref{s:LOHOL}), respectively.
\item
	For each of the presented systems, we give an interpretation of formulas and sequents into an underlying semantic universe.
\item
	We provide a series of examples, dealing with properties like continuity and differentiability, showing how they can be handled in the logical systems. These are spread out in Section~\ref{s:OHOL} and Section~\ref{s:LOHOL}.
\end{itemize}

\section{Reasoning about First-Order Properties, Compositionally}
\label{s:Reasoning}	

Before moving to the technical development of open higher-order logic, we informally 
describe the kind of problems such a logic is meant to solve. 
We do so by means of an example that higher-order logic is \emph{in principle} capable of dealing with, but doing so is complex, at least if one wants to proceed compositionally:
continuity in the presence of higher-order functions.

%

Let us consider a simply typed $\lambda$-calculus extended with a base type $\type{Real}$ for real numbers, as well as with constants and symbols for functions of first-order type, not necessarily interpreted as continuous functions. It is clear that a term $t$ of type $\type{Real} \rightarrow \type{Real}$, in a situation like the one just mentioned, computes a function whose continuity properties depend on how it is constructed and on which constants occur within it. In UHOL \cite{Aguirre2020:RelationalLogics}, the unary variant of higher order logic (HOL), the fact that such a term $t$ actually computes a continuous function can be expressed as the formula $\cont(t)$,
where $\cont$ is a unary predicate on terms of type $\type{Real}\rightarrow\type{Real}$. Clearly, this predicate's properties, at least some of them, should be captured by way of logical formulas, which become axioms in the underlying formal system. We would thus have, e.g., an axiom about stability of continuity by composition, namely the following formula:
$$\forall p,q : \type{Real} \rightarrow \type{Real}, (\cont(p) \land \cont(q)) \Rightarrow \cont(\la x. q~(p~x))$$
Now, let $t$ be the term $\la x.h~(g~(f~x))$, where $f,g,h$ are all constants of type $\type{Real}\rightarrow\type{Real}$ interpreted as continuous functions. We would like to prove within UHOL, possibly in a compositional way, that $t$ --- itself a term of type  $\type{Real}\rightarrow\type{Real}$ --- is continuous too. By design, UHOL's rules --- and thus UHOL's proofs ---
follow the term structure target formulas refer to; and in the case of $\lambda$-abstractions, 
the corresponding rule can be read as follows: ``if, whenever the argument $x$ satisfies a precondition $\phi$, the body $u$ satisfies the postcondition $\psi$, then $\lambda x.u$ satisfies the formula $\forall x. \phi \Rightarrow \psi$.''
When it comes to $\cont(t)$, this means we need a postcondition satisfied by the term $h~(g~(f~x))$ \emph{for every} $x$. But continuity on $x$ cannot be defined looking only at (the semantics of) $h~(g~(f~x))$, namely at a single real number, and thus it is not clear how one should proceed. 

Open higher-order logic (OHOL), instead, considers open objects as first-class citizens, without altering the rest of the framework in any other way, so still decomposing terms in the style of logical relations. In other words, in open higher-order logic we can reason on the open term $h~(g~(f~x))$ seeing it \emph{as a function} of type $\type{Real}\rightarrow\type{Real}$. Formally, in this case, we consider open terms for the context $\Theta = x : \type{Real}$. The proof can then proceed as follows:

\begin{prooftree}
	\AXC{$\Gamma \mid \Psi \pT f: \type{Real} \rightarrow \type{Real} \set{\forall y: \type{Real}, \cont(y) \Rightarrow \cont(\rr~y)}$}
	\AXC{$\Gamma \mid \Psi \pT (g~(h~x)) : \type{Real} \set{\cont(\rr)}$}
	\BIC{$\Gamma \mid \Psi \pT f~(g~(h~x)) : \type{Real} \set{\cont(\rr)}$}
\end{prooftree}
with the following tree for the second premise.
\begin{prooftree}
	\AXC{$\Gamma \mid \Psi \pT g: \type{Real} \rightarrow \type{Real} \set{\forall y: \type{Real}, \cont(y) \Rightarrow \cont(\rr~y)}$}
	\AXC{$\Gamma \mid \Psi \pT (h~x): \type{Real} \set{\cont(\rr)}$}
	\BIC{$\Gamma \mid \Psi \pT (g~(h~x)) : \type{Real} \set{\cont(\rr)}$}
\end{prooftree}

The application rule that we use for this proof is basically an elimination of the implication constructor for the target formula. The derivation for the functions $f$ and $g$ would then be obtained using an axiom assessing that continuity composes, whereas the continuity of $h~x$ would then follow by hypothesis, as we assumed that all three functions were continuous and thus, in particular, we have $\cont(h~x)$.

The above example relies on continuity to show a limitation of (U)HOL. 
Such a limitation, however, is not specific to continuity, and it
virtually affects any first order property. 
In this paper, we shall study another interesting example involving a first-order property, 
this time at a relational level: correctness of a state-of-the-art algorithm for automatic differentiation. 
In fact, an algorithm for forward mode differentiation of simply-typed terms has been recently proved correct by way of open logical relations~\cite{Barthe2020:OpenLogicalRelations}. Formalizing the aforementioned correctness proof in higher-order logic, however, would pose problems of exactly the same kind as those we described above, since derivability and (automatic) differentiation only make sense when spelled out on functions. In Section~\ref{s:Diff} , 
we show how to solve this problem by a relational version of OHOL capable of 
dealing both with (open) terms $t$ \emph{and} with their derivatives $D(t)$.  

\section{Preliminaries}
\label{s:Prelim}

The (higher-order) logic we deal with in this work is, in its bare essence, a formal 
framework to prove properties about higher-order programs. Consequently, 
a precise definition of such a logic requires a formal definition of what a program is. 
Here, we consider a $\la$-calculus with base types, constants, and functions \cite{Barthe2020:OpenLogicalRelations}. 
We write $\mathcal{C}$ and $\mathcal{F}$ for the collections of constants and 
symbols for first-order functions upon which terms are defined. 
The syntax and statics of the resulting calculus, that we denote by $\LaCF$, are given in Figure~\ref{f:calculusdescription}. 
The metavariable $\B$ ranges over \emph{base types} 
 (we assume a fixed collection of base types as given), such as real numbers or booleans. Moreover, we assume each constant $c \in \mathcal{C}$ to inhabit a base type $\B$ (notation $c : \B \in \mathcal{C}$) and each function $f \in \mathcal{F}$ to inhabit a 
 function type $\B_1 \times \cdots \times \B_m \rightarrow \B$ (notation $f : \B_1 \times \cdots \times \B_m \rightarrow \B \in \mathcal{F}$). We oftentimes employ the notation $\seq{\B}$ to denote 
a tuple $\B_1 \times \cdots \times \B_m$ when $m$ is clear from the context. This notation generalizes to other objects, e.g. terms, types, and typing contexts.

\begin{figure}
	\begin{framed}
		\begin{center}		
		$t,u ::= x \midd \und{c} \midd \und{f} \midd \la y. t \midd t \app u \midd \pair{t,u} \midd \fst(t) \midd \snd(t)$
		\qquad 
		$\tau,\sigma ::= \B \midd \tau \times \sigma \midd \tau \rightarrow \sigma$
		\\
			\vvskip
			\AXC{}
			\UIC{$\Gamma, x : \tau \p x : \tau$}
			\DP
			\qquad 
			\AXC{$c : \B \in \mathcal{C}$}
			\UIC{$\Gamma \p \und{c} : \B$}
			\DP
			\qquad
			\AXC{$f : \seq{\B} \rightarrow \B \in \mathcal{F}$}
			\UIC{$\Gamma \p \und{f} : \seq{\B} \rightarrow \B$}
			\DP
			\qquad 
			\AXC{$\Gamma, x : \tau \p t : \sigma $}
			\UIC{$\Gamma \p \la x. t : \tau \rightarrow \sigma$}
			\DP
			\\ 
			\vvskip 
			\AXC{$\Gamma \p t : \tau \rightarrow \sigma$}
			\AXC{$\Gamma \p u : \tau$}
			\BIC{$\Gamma \p t \app u : \sigma$} 
			\DP 
			\qquad 
			\AXC{$\Gamma \p t : \tau_1$}
			\AXC{$\Gamma \p u : \tau_2$}
			\BIC{$\Gamma \p \pair{t,u} : \tau_1 \times \tau_2$}
			\DP
			\qquad
			\AXC{$\Gamma \p t : \tau_1 \times \tau_2$}
			\UIC{$\Gamma \p \pi_i(t) : \tau_i$}
			\DP
		\end{center}		
	\end{framed}
	\caption{Syntax and Static Semantics for $\LaCF$}
	\label{f:calculusdescription}
\end{figure} 

Finally, we endow  $\LaCF$ with a standard set-theoretic 
denotational semantics in the usual way. Accordingly, each type $\tau$ is interpreted 
as a set 
$\sem{\tau}$, and any derivable typing judgment $\Gamma \p t : \tau$ is 
interpreted as a function $\sem{\Gamma \p t : \tau} : \sem{\Gamma} \rightarrow \sem{\tau}$, where $\sem{\Gamma} = \prod_{y:\sigma \in \Gamma} \sem{\sigma}$.
\longversion{This is described in Figure~\ref{f:calculusdenotational}

\begin{figure}
	\begin{framed}
		\begin{center}
			Suppose given for each base type $\B$ an interpretation $\sem{\B}$ (for example, the type for reals could be given as an interpretation the actual set of real numbers). Suppose also that for each constant $c : \B \in \mathcal{C}$, we have an interpretation $\sem{c} \in \sem{\B}$ and for each function $f : \seq{\B} \rightarrow \B \in \mathcal{F}$, we have an interpretation $\sem{f} :  \sem{\B}^{\sem{\seq{\B}}}$.
			\\ 
			\vvskip 
			Then, the interpretation of types is defined by an object in the category of $\catname{Set}$: \\
			\vvskip  
			$\tau_1 \times \tau_2 = \sem{\tau_1} \times \sem{\tau_2} \qquad \tau_1 \rightarrow \tau_2 = \sem{\tau_2}^{\sem{\tau_1}}$ 
			\\
			\vvskip 
			And, a typing derivation $\Gamma \p t : \tau$ is interpreted as a map $\sem{\Gamma \p t : \tau} : \sem{\Gamma} \rightarrow \sem{\tau}$, this is defined by induction on the type derivation with: 
			\\ 
			\vvskip 
			$\sem{\Gamma, x : \tau \p x : \tau}(\seq{y},y) = y \qquad \sem{\Gamma \p \und{c} : \B}(\seq{y}) = \sem{c} \qquad \sem{\Gamma \p \und{f} : \seq{\B} \rightarrow \B}(\seq{y}) = \sem{f}$
			\\ 
			\vvskip 
			$\sem{\Gamma \p \lambda x. t : \tau_1 \rightarrow \tau_2}(\seq{y}) = \mathtt{fun}~(y : \sem{\tau_1}) \mapsto \sem{\Gamma, x: \tau_1 \p t : \tau_2}(\seq{y},y)$
			\\ 
			\vvskip
			$\sem{\Gamma \p t_1 \app t_2 : \tau_2}(\seq{y}) = (\sem{\Gamma \p t_1 : \tau_1 \rightarrow \tau_2}(\seq{y}))(\sem{\Gamma \p t_2 : \tau_1}(\seq{y}))$
			\\ 
			\vvskip 
			$\sem{\Gamma \p \pair{t_1,t_2} : \tau_1 \times \tau_2}(\seq{y}) = (\sem{\Gamma \p t_1 : \tau_1}(\seq{y}),\sem{\Gamma \p t_2 : \tau_2}(\seq{y})) $
			\\
			\vvskip 
			$\sem{\Gamma \p \pi_i(t) : \tau_i}(\seq{y}) = \pi_i(\sem{\Gamma \p t : \tau_1 \times \tau_2}(\seq{y}))$ 
			
		\end{center}
	\end{framed}
	\caption{Denotational Semantics for $\LaCF$}
	\label{f:calculusdenotational}
\end{figure}
}

\subsection{Higher-Order Logic}
Having defined $\LaCF$, we now recall basic definitions of \emph{Higher-Order Logic} \cite{Aguirre2020:RelationalLogics}, 
the logic we will build upon.
\begin{definition}
The syntax of HOL formulas is given by the following grammar:
\begin{align*}
	\phi ::= P(t_1,\dots,t_m) \midd \top \midd \bot \midd \phi \land \phi \midd \phi \lor \phi \midd \phi \Rightarrow \phi \midd \forall y:\tau. \phi \midd \exists y: \tau. \phi.
\end{align*}
\end{definition}
HOL formulas  are defined starting from a given collection of atomic predicates on $\LaCF$ terms, 
from which more complex formulas are then constructed relying on connectives. Each predicate $P$ comes with an \emph{arity} $(\tau_1,\dots,\tau_m)$ stating that in an atomic formula $P(t_1,\dots,t_m)$ each term $t_i$ must have type $\tau_i$. Moreover, notice that variables introduced by quantifiers are (typed) term variables, meaning that they can occur in terms themselves occurring in atomic formulas. 
 This intuitive description is formalized by means of well-typing judgments of the form  
 $\Gamma \p \phi$ whose defining rules are as expected. Due to space constraints, we 
 omit the formal definition of such rules and address the reader to one 
 of the many papers on the subject \cite{Aguirre2017:RelationalLogic,Aguirre2020:RelationalLogics}. 
 As it is customary, we also assume an equality predicate to be available for \emph{all} types. 

HOL inherits the set-theoretical semantics of $\LaCF$: 
 given an interpretation $\sem{P} \subseteq \sem{\tau_1} \times \cdots \times \sem{\tau_m}$ for each predicate $P$ of arity $(\tau_1,\dots,\tau_m)$,
well-typed assertions $\Gamma \p \phi$ are interpreted as subsets $\sem{\Gamma \p \phi} \subseteq \sem{\Gamma}$.
Intuitively, this semantics is defined for non-atomic formulas by the standard interpretation of boolean constructors and quantification, 
whereas we define the interpretation $\sem{P(t_1,\dots,t_m)}$ of 
an atomic formula $P(t_1,\dots,t_m)$ as the set of elements $\gamma \in \sem{\Gamma}$ such that $(\sem{\Gamma \p t_i : \tau_i}(\gamma))_{1 \le i \le m} \in \sem{P}$.  

The real power of HOL is not its set-theoretic semantics, but its proof theory. 
In fact, HOL comes with a proof system that we recall here in a sequent calculus-style. 
Such a system employs judgments of the form $\Gamma \mid \Psi \p \phi$, where $\Psi$ is a set of well-typed assertions, and $\phi$ is a well-typed assertion. A valid judgment $\Gamma \mid \Psi \p \phi$ attests that in the typing context $\Gamma$ and under the assumptions in $\Psi$, $\phi$ is true. 
Accordingly, predicates $P$ must come with axioms defining their (logical) meaning, as there is no rule for arbitrary predicates. 
We recall some of the most important rules in Figure~\ref{f:HOLjudgments}, other rules being the usual inference rules for logical constructors (we write $=_{(\rightarrow)}$ for the smallest equivalence relation including $\rightarrow$, the reduction relation on $\LaCF$, which can be defined as expected). Note that in addition to the logical rules, there are rules specific to the equality predicate allowing, in particular, term substitutions. 

\begin{figure}
	\begin{framed}
			\begin{center} 
			\AXC{}
			\UIC{$\Gamma \mid \Psi, \phi \p \phi $}
			\DP 
			\qquad 
			\AXC{$\Gamma \p t_i : \tau$}
			\AXC{$t_1 =_{(\rightarrow)} t_2$}
			\BIC{$\Gamma \mid \Psi \p  (t_1 = t_2)$}
			\DP
			\qquad  
			\AXC{$\Gamma \mid \Psi \p  \phi \sub{y}{t_1}$}
			\AXC{$\Gamma \mid \Psi \p  (t_1 = t_2) $}
			\BIC{$\Gamma \mid \Psi \p  \phi \sub{y}{t_2}$}
			\DP 
			\\
			\vvskip 
			\AXC{$\Gamma, y : \tau \mid \Psi \p \phi$}
			\UIC{$\Gamma \mid \Psi \p \forall y : \tau. \phi $}
			\DP 
			\qquad  
			\AXC{$\Gamma \mid \Psi \p \forall y : \tau. \phi $}
			\AXC{$\Gamma \p t : \tau$}
			\BIC{$\Gamma \mid \Psi \p \phi \sub{y}{t} $}
			\DP 
		\end{center}
	\end{framed}
	\caption{Selected Rules of HOL}
	\label{f:HOLjudgments}	
\end{figure}

\subsection{Unary HOL}
In practical examples, especially in presence of unary predicates, (the proof system of) HOL may be difficult to use, its rules being ultimately formula-directed. In those cases, 
it is desirable to have a system (whose rules are) directed by the structure 
of the terms involved.
Moving from this observation, Aguirre et al. 
\cite{Aguirre2020:RelationalLogics,Aguirre2017:RelationalLogic} have developed
term-directed proof systems for relational higher-order logics. We recall how such systems work, focusing on the unary case only. 

\begin{definition} 
 \emph{Unary} higher-order logic (UHOL) is an inference system based on judgments of the form 
 $\Gamma \mid \Psi \p t : \tau \set{\phi}$, where $\Gamma$ and $t$ are as usual, 
 $\Psi$ is a set of HOL formulas, and $\phi$ is a HOL formula with a distinguished free variable 
 $\rr$ not appearing in $\Gamma$ (i.e. $\rr$ acts as a placeholder for $t$ in the \emph{target formula} $\phi$). 
 The defining rules of UHOL are given in Figure~\ref{f:UHOLjudgments}.
 \end{definition}
 
 A UHOL judgment $\Gamma \mid \Psi \p t : \tau \set{\phi}$ attests that $t$ has type $\tau$ 
 in the typing context $\Gamma$, and that the formula $\phi\sub{\rr}{t}$ is true 
 under the assumptions in $\Psi$. 
To prove such judgments, we rely on term-directed rules. 
In particular, the first three rules in Figure~\ref{f:UHOLjudgments} state that for base terms, satisfiability of the target formula must be verified in the HOL judgment system. Other rules reshape the target formula according to the structure of $t$. For example, in the $\la$-abstraction rule the target formula expresses that if an argument satisfies a precondition $\psi$, then the application satisfies a postcondition $\phi$. This can be verified assuming $\psi$ and trying to prove the target formula $\phi$ for the body of the $\la$-abstraction. Finally, the last rule, that does not depend on the structure of $t$, allows us to use HOL 
judgments to weaken target formulas. 

\begin{figure}
	\begin{framed}
		\begin{center}
			\AXC{$\Gamma, y : \tau \mid \Psi \p \phi\sub{\rr}{y}$}
			\UIC{$\Gamma, y : \tau \mid \Psi \p y : \tau \set{\phi}$}
			\DP
			\qquad 
			\AXC{$\Gamma \mid \Psi \p \phi\sub{\rr}{\und{c}}$}
			\UIC{$\Gamma \mid \Psi \p \und{c} : \B \set{\phi}$}
			\DP
			\\ 
			\vvskip 
			\AXC{$\Gamma \mid \Psi \p \phi\sub{\rr}{\und{f}}$}
			\UIC{$\Gamma \mid \Psi \p \und{f} : \seq{\B} \rightarrow \B \set{\phi}$}
			\DP
			\qquad 
			\AXC{$\Gamma, y : \tau \mid \Psi,\psi \p t : \sigma \set{\phi}$}
			\UIC{$\Gamma \mid \Psi \p \la y. t : \tau \rightarrow \sigma \set{\forall y. \psi \Rightarrow \phi\sub{\rr}{\rr \app y}}$}
			\DP
			\\ 
			\vvskip
			\AXC{$\Gamma \mid \Psi \p t : \tau \rightarrow \sigma \set{\forall y. \psi\sub{\rr}{y} \Rightarrow \phi\sub{\rr}{\rr \app y}}$}
			\AXC{$\Gamma \mid \Psi \p u : \tau \set{\psi}$}
			\BIC{$\Gamma \mid \Psi \p t \app u : \sigma \set{\phi\sub{y}{u}}$}
			\DP
			\\
			\vvskip
			\AXC{$\Gamma \mid \Psi \p t_i : \tau_i \set{\phi_i}$}
			\AXC{$\Gamma \mid \Psi \p \forall y,z. \phi_1\sub{\rr}{y} \land \phi_2\sub{\rr}{z} \Rightarrow \phi\sub{\rr}{\pair{y,z}}$}
			\BIC{$\Gamma \mid \Psi \p \pair{t_1,t_2} : \tau_1 \times \tau_2 \set{\phi}$}
			\DP
			\\ 
			\vvskip
			\AXC{$\Gamma \mid \Psi \p t : \tau_1 \times \tau_2 \set{\phi\sub{\rr}{\pi_i(\rr)}}$}
			\UIC{$\Gamma \mid \Psi \p \pi_i(t) : \tau_i \set{\phi}$}
			\DP
			\qquad
			\AXC{$\Gamma \mid \Psi \p t : \tau \set{\psi}$}
			\AXC{$\Gamma \mid \Psi \p \psi\sub{\rr}{t} \Rightarrow \phi\sub{\rr}{t}$}
			\BIC{$\Gamma \mid \Psi \p t : \tau \set{\phi}$}
			\DP
		\end{center}
	\end{framed}
	\caption{Rules of UHOL}
	\label{f:UHOLjudgments}
\end{figure}

Aguirre et al.'s results \cite{Aguirre2020:RelationalLogics,Aguirre2017:RelationalLogic}  show that 
HOL and UHOL are equivalent --- i.e. $\Gamma \mid \Psi \p t : \sigma \set{\phi}$ if and only if $\Gamma \mid \Psi \p \phi\sub{\rr}{t}$ --- and that they are both sound in regard to the previously 
introduced set-theoretic semantics: if $\Gamma \mid \Psi \p \phi$ is valid, then $\sem{\Gamma \p (\bigwedge_{\psi \in \Psi} \psi)} \subseteq \sem{\Gamma \p \phi}$. 
But even if equivalent,  HOL and UHOL show several differences of usage in practice. 

In the HOL system, it is difficult to know when to use a term substitution to simplify a term, as well as 
when to apply the cut rule. Consequently, the system turns out to work well on judgments speaking about simple enough terms, 
but it may be difficult to use when more complex terms are involved. 
In UHOL, instead, the last rule of Figure~\ref{f:UHOLjudgments} (which is crucial to guarantee  expressiveness of the logic) turns out to be problematic for automation, and it is thus desirable to avoid its usage as much as possible. This makes the proof system of UHOL effective when applied to formulas with possibly complex terms, 
but only if there is little or no need to adjust the target formula (i.e. to use the last rule in  Figure~\ref{f:UHOLjudgments}).
UHOL also relies on HOL to prove properties of base terms, so that it inherits the same weaknesses of 
the latter 
Finally, the system is designed to work on terms and formulas whose shape matches the semantic 
behavior of the terms they refer to. In absence of such a correspondence, proofs become 
remarkably difficult. 
The typical example of such a behavior is the one we described in Section~\ref{s:Reasoning}: when a formula for a $\la$-abstraction focuses on the whole function --- and not on the actual application of this function to an argument --- the rule for $\la$-abstraction cannot be used. This motivates our work and the definition of Open HOL, as currently neither HOL nor UHOL can easily cope with first-order properties of programs.

\section{Open Higher-Order Logic}
\label{s:OHOL}

In this section, we introduce \emph{Open} HOL (OHOL, for short), 
a higher-order logic designed to natively deal with \emph{first-order properties} 
of programs. To do so, we follow the idea behind open logical relations~\cite{Barthe2020:OpenLogicalRelations} where, in the base case of the (recursive) 
definition of a logical relation, \emph{closed} terms 
of a ground type $\B$ --- which, semantically, correspond to elements in $\sem{\B}$ ---
are replaced by \emph{open} terms having free variables in a 
fixed typing context $\Theta$ --- so that the semantics of such terms 
is given by \emph{functions} from $\sem{\Theta}$ to 
$\sem{\B}$, which are then required to satisfy the first-order property of interest.

Let us fix a typing context $\Theta = \xx_1 : \B_1, \dots, \xx_n : \B_n$. For the sake of clarity, we consider variables $\xx_1, \dots, \xx_n$ as disjoint from the usual term variables, which we denote by $y, y_1, \dots$. 
OHOL's formulas are defined by the following grammar:
\begin{align*}
	\phi ::= P^{\Theta}(t_1,\dots,t_m) \midd \top \midd \bot \midd \phi \land \phi \midd \phi \lor \phi \midd \phi \Rightarrow \phi \midd \forall y:\tau. \phi \midd \exists y: \tau. \phi.  
\end{align*}
Notice that there is only one difference between HOL's and OHOL's formulas, namely 
atomic predicates. In OHOL, in fact, atomic predicates $P^{\Theta}$ 
are parametrized by the typing context $\Theta$, with the intended meaning that arguments 
of $P^{\Theta}$ may have free variables in $\Theta$. That is, 
if a predicate $P^{\Theta}$ has arity $(\tau_1,\dots,\tau_n)$, then in an atomic formula $P^{\Theta}(t_1,\dots,t_m)$ we require 
each term $t_i$ to have type $\tau_i$ in the typing context $\Theta$. 
Formally, this implies that to derive well-typed OHOL's judgments  $\Gamma \pT \phi$,
we shall rely on the following rule for atomic predicates: 
\begin{prooftree}
	\AXC{$P^{\Theta}$ has arity $(\tau_1,\dots,\tau_m)$}
	\AXC{$\forall 1 \le i \le m,~ \Gamma, \Theta \p t_i : \tau_i$}
	\BIC{$\Gamma \pT P^{\Theta}(t_1,\dots,t_m)$}
\end{prooftree}

Let us now see how the admittedly minor differences between the defining grammars
 of HOL and OHOL formulas (and, consequently, in their judgment rules) 
 impact on the logics themselves. 
 We begin with set-theoretic semantics, where we see that OHOL indeed interprets
  a term  $t$ of type $\tau$ with free variables in $\Theta$ as a function from $\sem{\Theta}$ to $\sem{\tau}$.
More generally, we interpret well-typed assertions $\Gamma \pT \phi$ as subsets $\sem{\Gamma \pT \phi} \subseteq \sem{\Gamma^\Theta}$ with $\sem{\Gamma^\Theta} \cong \prod_{y : \tau \in \Gamma} \sem{\tau}^{\sem{\Theta}}$. 
For readability, it is convenient to introduce the following notation: given $f \in B^{A \times C}$ and $g \in A^C$, we define $f \Comp g \in B^{C}$ by $(f \Comp g)(x) = f(g(x),x)$. 

\begin{definition}
\label{def:OHOL-set-theoretic-semantics}
Given an interpretation 
$\sem{P^{\Theta}} \subseteq \prod_{1 \le i \le m}  \sem{\tau_i}^{\sem{\Theta}}$ 
of predicates $P^{\Theta}$ of arity $(\tau_1,\dots,\tau_m)$, 
the semantics $\sem{\Gamma \pT P^{\Theta}(t_1,\dots,t_m)} \subseteq 
\sem{\Gamma}^{\sem{\Theta}}$ 
of the judgment $\Gamma \pT P^{\Theta}(t_1,\dots,t_m)$ is defined as follows:
\begin{center}
$
\sem{\Gamma \pT P^{\Theta}(t_1,\dots,t_m)} = \{ \seq{y} \in \sem{\Gamma^\Theta} \midd \prod_{1 \le i \le m}(\sem{\Gamma, \Theta \p t_i : \tau_i} \Comp \seq{y}) \in \sem{P^{\Theta}}\}.
$
\end{center}
We define the semantics $\sem{\Gamma \pT \phi} \subseteq \sem{\Gamma}^{\sem{\Theta}}$ 
of a well-typed judgment 
$\Gamma \pT \phi$ by 
inductively extending $\sem{\Gamma \pT P^{\Theta}(t_1,\dots,t_m)}$ 
in the usual way. For instance: 
\begin{align*}
\sem{\Gamma \pT \phi \lor \psi} 
&= \sem{\Gamma \pT \phi} \cup \sem{\Gamma \pT \psi};
\\
\sem{\Gamma \pT \forall y:\tau. \phi} 
&= \{\seq{y} \in \sem{\Gamma^{\Theta}} \midd \forall y \in \sem{\tau}^{\sem{\Theta}}, (\seq{y},y) \in \sem{\Gamma, y : \tau \pT \phi}\}.
\end{align*}
\end{definition}

Notice that even if OHOL's formulas are essentially those of HOL, 
Definition~\ref{def:OHOL-set-theoretic-semantics} interprets 
terms and variables of type $\tau$ as function from $\sem{\Theta}$ to $\sem{\tau}$. 
In particular, notice that in the formula $\forall y : \tau. \phi$ the variable $y$ has type $\tau$ in the underlying context, but it is interpreted as a function in $\sem{\tau}^{\sem{\Theta}}$.

Having defined the set-theoretic semantics of OHOL, we move 
to its proof system. We consider judgments of the form $\Gamma \mid \Psi \pT \phi$ 
and  adjust the judgment system of HOL (Figure~\ref{f:HOLjudgments}) 
by adding the typing context $\Theta$ to all premises of a type derivation of $\LaCF$ 
terms. For example the rule
\[
\vcenter{
\infer{\Gamma \mid \Psi \p \phi\sub{y}{t}}
{\Gamma \mid \Psi \p \forall y : \tau. \phi 
&\Gamma \p t : \tau}
}
\quad
\text{is replaced by}
\quad
\vcenter{
\infer{\Gamma \mid \Psi \pT \phi\sub{y}{t}}
{\Gamma \mid \Psi \pT \forall y : \tau. \phi
&\Gamma, \Theta \p t : \tau}
}
\]

\begin{proposition}
OHOL is sound with respect to the set-theoretic semantics. That is, 
if $\Gamma \mid \Psi \pT \phi$ is derivable, then 
$\sem{\Gamma \pT \bigwedge_{\psi \in \Psi} \psi} 
\subseteq \sem{\Gamma \pT \phi}$.
\end{proposition} 
\begin{proof}
By induction on $\Gamma \mid \Psi \pT \phi$ proceeding as in 
the corresponding proof for HOL.
\end{proof}

\subsection{Unary OHOL}
The OHOL proof system, being ultimately defined upon the one for HOL, 
shares (some) strengths and weaknesses with the latter. In particular, 
as for HOL, the formula-directed nature of OHOL rules makes OHOL difficult 
to use in presence of complex terms.
To overcome this problem, we proceed as in the design of UHOL 
and introduce a new judgment system, which we dub \emph{Unary Open HOL} (UOHOL, for short), whose proof rules are term-directed. 
Formally, UOHOL has judgments of the form $\Gamma \mid \Psi \pT t : \tau \set{\phi}$, where $\Psi$ is a set of OHOL formulas, $\phi$ is a OHOL formula with a distinguished variable $\rr$, and $t$ is a $\LaCF$ term. A judgment $\Gamma \mid \Psi \pT t : \tau \set{\phi}$ states that $\Gamma, \Theta \p t: \tau$ and that, under the assumptions in $\Psi$, $\phi \sub{\rr}{t}$ is satisfied. UOHOL rules consist of the previously mentioned rules of Figure~\ref{f:UHOLjudgments}, with one additional rule:
\begin{prooftree}
	\AXC{$(\xx_i : \B_i) \in \Theta$}
	\AXC{$\Gamma \mid \Psi \pT \phi\sub{\rr}{\xx_i}$}
	\BIC{$\Gamma \mid \Psi \pT \xx_i : \B_i \set{\phi}$}
\end{prooftree}
Such a rule corresponds to the axiom rule in Figure~\ref{f:UHOLjudgments}, but for variables in $\Theta$. 
Finally, following the case of HOL and UHOL, we prove equivalence of  OHOL and UOHOL.

\begin{proposition} 
UOHOL and OHOL are equivalent, i.e. 
$\Gamma \mid \Psi \pT t : \tau \set{\phi}$ if and only if $\Gamma \mid \Psi \pT \phi \sub{\rr}{t}$.
\end{proposition}

Compared to UHOL, we see that UOHOL allows for more satisfactory proofs of higher-order properties, as mentioned in Section~\ref{s:Reasoning}. To see that, let $\phi$ 
be the following OHOL formula:
$\forall y : \tau \rightarrow \sigma, \forall z : \tau, P^{\Theta}(y~z).$
It is easy to see that $\phi$ could be written as a HOL formula too: 
for, let us consider the formula $\psi$ below, with $T$ being product of 
types in $\Theta$:
$$\forall y : T \rightarrow (\tau \rightarrow \sigma), \forall z : (T \rightarrow \tau), P(\la x. (y~x)~(z~x)).$$
Compared to $\psi$, the formula $\phi$ is considerably simpler, and thus it facilitates derivation in OHOL or UOHOL. Moreover, 
it is precisely the simpler (logical) structure of $\phi$ that allows us to 
deal with  first-order properties smoothly, in contrast to the 
previously mentioned UHOL's weaknesses. 
This is the main point of OHOL and UOHOL. We do not claim any increase 
of expressiveness over the standard HOL. What we claim, instead, is that OHOL and 
UOHOL constitute a way to simplify the process of proving some formulas of HOL that cannot be readily proved valid in HOL or UHOL.

\subsection{OHOL with Multiple Domains and ROHOL}
\label{s:Diff}

The discussion at the end of Section~\ref{s:Prelim} on the drawbacks and benefits of HOL and UHOL remains relevant for OHOL and UOHOL. 
Moreover, even if OHOL solves some problems suffered by HOL when applied to 
first-order properties, it does \emph{not} solve \emph{all} of them. 
In particular, OHOL focuses on a single domain --- the typing context $\Theta$ ---
that cannot be changed in a formula. Consequently, all terms in a given formula 
are relative to the \emph{same} domain $\Theta$. As we are going to see, however,  
it is sometimes useful to allow different terms to be associated to different typing contexts 
within the same formula. Here, we present a simple generalization of OHOL designed to 
tackle this problem, showing also its usefulness for the relational version of UOHOL, that we briefly sketch below.

\subsubsection{Handling Multiples Domains}

Let us begin with OHOL with multiple domains of definition. 
\shortversion{
Due to space constraints, the formal description of the logic has 
to be omitted (but see \cite{TechReport} for details). Nonetheless, we can still outline its main features here, although at an informal level only.}
In OHOL with multiple domains of definition, we associate to each variable and term a unique typing context, but contrary to OHOL, we allow different terms to be associated with \emph{distinct} typing contexts. 
In particular, we assign to variables not only types but 
also typing contexts, which now play the role of kinds. Consequently, we have typed variables
$y: \tau :: \Theta$ stating that $y$ has type $\tau$, and $\tau$ has kind $\Theta$. 
Semantically, we let kinds specify the domains of the semantic interpretation of variables:
given typing contexts $\Theta_1$ and $\Theta_2$, the variables 
$y_1: \tau_1 :: \Theta_1$ and $y_2: \tau_2 :: \Theta_2$ are interpreted 
as functions in $\sem{\tau_1}^{\sem{\Theta_1}}$ and $\sem{\tau_2}^{\sem{\Theta_2}}$, respectively 
(recall that in OHOL each variable $y:\tau$ is interpreted as a function 
$\sem{\tau}^{\sem{\Theta}}$, with $\Theta$ \emph{fixed}).

The semantic interpretation for formulas using kinds is a rather straightforward generalization of the semantics in Definition~\ref{def:OHOL-set-theoretic-semantics}: 
we only need to modify the notion of arity for predicates, which are now of 
the form $(\tau_1 :: \Theta_1, \tau_2 :: \Theta_2, \dots, \tau_m :: \Theta_m)$, 
with the intended meaning that type $\tau_i$ has kind $\Theta_i$. 
Therefore, a predicate of arity 
$(\tau_1 :: \Theta_1, \tau_2 :: \Theta_2, \dots, \tau_m :: \Theta_m)$
is now interpreted as a subset of $\prod_{1 \leq i\leq m} \sem{\tau_i}^{\sem{\Theta_i}}$.
\longversion{
	Formally, the semantics is described in Figure~\ref{f:OHOLSemanticsKinds}. First, we define the semantics of contexts by:
	$$ \sem{\cdot} = 1 \qquad \sem{\Gamma, y : \tau :: \Theta} = \sem{\Gamma} \times  \sem{\tau}^{\sem{\Theta}}) $$
	We then give the following notation to separate contexts. If $\Gamma$ is a context and $\Theta$ a kind, we define $\Gamma(\Theta)$ as the set of assertions $y : \tau$ such that $y : \tau :: \Theta \in \Gamma$. Given a sequence of elements $\seq{y} \in \sem{\Gamma}$, we define $\seq{y}(\Theta)$ as the subsequence of elements corresponding to variables of kind $\Theta$ in $\Gamma$. By definition, we then have $\seq{y}(\Theta) \in \sem{\Gamma(\Theta)}^{\sem{\Theta}}$. 
\begin{figure}
	\begin{framed}
		With an interpretation $\sem{P^{\Theta}} \subseteq \prod\limits_{1 \le i \le m} \sem{\tau_i}^{\sem{\Theta_i}}$ for each predicate $P^{\Theta}$ of arity $(\tau_1 :: \Theta_1,\dots,\tau_m :: \Theta_m)$, the semantics $\sem{\Gamma \p \phi} \subseteq  (\sem{\Gamma})$ is defined by induction on $\Gamma \p \phi$:
		\begin{itemize}
			\item $\sem{\Gamma \p P(t_1 :: \Theta_1,\dots,t_m :: \Theta_m)} = \Big\{ \seq{y} \in \sem{\Gamma} \midd \prod\limits_{1 \le i \le m}(\sem{\Gamma(\Theta_i), \Theta_i \p t_i : \tau_i} \Comp \seq{y}(\Theta_i)) \in \sem{P} \Big\}$
			\item 
			$\sem{\Gamma \p \top} = \sem{\Gamma}$
			\item 
			$\sem{\Gamma \p \bot} = \emptyset$
			\item 
			$\sem{\Gamma \p \phi_1 \land \phi_2} = \sem{\Gamma \p \phi_1} \cap \sem{\Gamma \p \phi_2}$
			\item 
			$\sem{\Gamma \p \phi_1 \lor \phi_2} = \sem{\Gamma \p \phi_1} \cup \sem{\Gamma \p \phi_2}$
			\item 
			$\sem{\Gamma \p \phi_1 \Rightarrow \phi_2} = (\sem{\Gamma} / \sem{\Gamma \p \phi_1}) \cup \sem{\Gamma \p \phi_2}$
			\item		
			$\sem{\Gamma \pT \forall y:\tau :: \Theta. \phi} = \Big\{\seq{y} \in \sem{\Gamma} \midd \forall y \in (\sem{\Theta} \Rightarrow \sem{\tau}), (\seq{y},y) \in \sem{\Gamma, y : \tau \pT \phi} \Big\}$
			\item 
			$\sem{\Gamma \pT \exists y:\tau :: \Theta. \phi} = \Big\{\seq{y} \in \sem{\Gamma} \midd \exists y \in (\sem{\Theta} \Rightarrow \sem{\tau}), (\seq{y},y) \in \sem{\Gamma, y : \tau \pT \phi} \Big\}$
		\end{itemize}
	\end{framed}
	\caption{Semantics of OHOL}
	\label{f:OHOLSemanticsKinds}
\end{figure}
}

As for the logical rules, the only relevant point is to specify a kind 
for each term and for each variable, and to impose that all variables in a term have the same kind. For example, we have the following rule for the introduction and elimination of the $\forall$ constructor:
\begin{center}
	\small 
	\AXC{$\Gamma, y : \tau :: \Theta \mid \Psi \p \phi$}
	\UIC{$\Gamma \mid \Psi \p \forall y : \tau :: \Theta. \phi $}
	\DP 
	\qquad 
	\AXC{$\Gamma\mid \Psi \p \forall y : \tau :: \Theta. \phi $}
	\AXC{$\Delta \p t : \tau$}
	\AXC{$\forall (z : \sigma) \in \Delta, (z : \sigma :: \Theta) \in \Gamma$}
	\TIC{$\Gamma \mid \Psi \p \phi \sub{y}{t} $}
	\DP 
\end{center}
\subsubsection{Going Relationally}
Even if theoretically fine, having access to different kinds is not that useful for
UOHOL. Indeed, in a UOHOL proof, the focus is on a \emph{single} term (as well as its subterms) only, 
so that if a term has kind $\Theta$, then we do not gain much from
using OHOL with multiple domains of definition.
The potential of such a logic, instead, becomes evident when we move 
 to \emph{relational} reasoning. 
Aguirre et al. \cite{Aguirre2017:RelationalLogic,Aguirre2020:RelationalLogics}
have introduced RHOL, a relational version of UHOL talking about \emph{pairs} of terms, 
rather than a single term in isolation. 
Because of the similarities between UHOL and UOHOL, the design of ROHOL from UOHOL can be done by mimicking the original construction by Aguirre et al. 
A RHOL judgment has the shape $\Gamma \mid \Psi \p t_1 : \tau_1 \sim t_2 : \tau_2 \set{\phi}$, where $\phi$ contains two special variables $\rr_1$ and $\rr_2$ 
acting as placeholders for $t_1$ and $t_2$, respectively. 
The intuitive meaning is that $t_1$ and $t_2$ have types $\tau_1$ and $t_2$, respectively, 
in the typing context $\Gamma$, and that  $\phi\sub{\rr_1}{t_1}\sub{\rr_2}{t_2}$ holds 
under the assumptions in $\Psi$.

 As an example, a RHOL rule for $\la$-abstraction is
\begin{center}
	\AXC{$\Gamma, y_1 : \sigma_1, y_2 : \sigma_2 \mid \Psi, \psi \p t_1 : \tau_1 \sim t_2 : \tau_2 \set{\phi}$}
	\UIC{$\Gamma \mid \Psi \p \la y_1. t_1 : \sigma_1 \rightarrow \tau_1 \sim \la y_2. t_2 : \sigma_2 \rightarrow \tau_2 \set{\forall y_1,y_2. \psi \Rightarrow \phi\sub{\rr_1}{\rr_1 \app y_1}\sub{\rr_2}{\rr_2 \app y_2}}$}
	\DP    
\end{center}
Such a rule can be seen as  two instances of the rule for $\lambda$-abstractions from UHOL merged together. 
In a similar fashion,  we can extend such a rule to our open setting with a fixed context $\Theta$. 
More interesting is the case of multiple domains of definition, 
where we can consider the first term having kind $\Theta_1$ and the second one having kind $\Theta_2$. This way, we obtain the following rule in what we may call ROHOL 
with multiple domains of definition:
\begin{center}
	\footnotesize
	\AXC{$\Gamma, y_1 : \sigma_1 	:: \Theta_1, y_2 : \sigma_2 :: \Theta_2 \mid \Psi, \psi \p t_1 : \tau_1 \sim t_2 : \tau_2 \set{\phi}$}
	\UIC{$\Gamma \mid \Psi \p \la y_1. t_1 : \sigma_1 \rightarrow \tau_1 \sim \la y_2. t_2 : \sigma_2 \rightarrow \tau_2 \set{\forall y_1 : \sigma_1 :: \Theta_1,y_2 : \sigma_2 :: \Theta_2. \psi \Rightarrow \phi\sub{\rr_1}{\rr_1 \app y_1}\sub{\rr_2}{\rr_2 \app y_2}}$}
	\DP    
\end{center}
Such a rule also allows us to see the benefits of having different kinds, as we are 
now allowed to relate functions with different initial domains. 
\shortversion{As before, space constraints force us to omit formal details 
(for which we refer to the long version of this paper \cite{TechReport}). 
Instead, we witness how our logic work by means of a nontrivial example.}

\longversion{We consider that $t_1$ has kind $\Theta_1$ and $t_2$ has kind $\Theta_2$ in a ROHOL judgment, so to simplify notations we will not rewrite those kind in each rules. This means in particular that the variable $\rr_1$ has kind $\Theta_1$ and that $\rr_2$ has kind $\Theta_2$. Some ROHOL rules are described in Figure~\ref{f:ROHOLjudgments}. Other important rules that we do not describe here as they are rules that follows the structure of only one of the two terms. The main idea is to apply a UOHOL rule on one of the two terms $t_1$ or $t_2$ and let the other one totally unchanged. Some examples can be found in \cite{Aguirre2020:RelationalLogics}. 

\begin{figure}
	\begin{framed}
		\begin{center}
			\AXC{$\Gamma(\Theta_1), \Theta_1 \p z_1 : \tau_1$}
			\AXC{$\Gamma(\Theta_2), \Theta_2 \p z_2 : \tau_2$}
			\AXC{$\Gamma \mid \Psi \p \phi\sub{\rr_1,\rr_2}{z_1,z_2}$}
			\TIC{$\Gamma_1 ; \Gamma_2 \mid \Psi \pTT z_1 : \tau_1 \sim z_2 : \tau_2 \set{\phi}$}
			\DP
			\\
			\vvskip
			\AXC{$\Gamma, y_1 : \sigma_1 :: \Theta_1, y_2 : \sigma_2 :: \Theta_2 \mid \Psi, \phi' \p t_1 : \tau_1 \sim t_2 : \tau_2 \set{\phi}$}
			\UIC{$\Gamma \mid \Psi \p \la y_1. t_1 : \sigma_1 \rightarrow \tau_1 \sim \la y_2. t_2 : \sigma_2 \rightarrow \tau_2 \set{\forall y_1,y_2. \phi' \Rightarrow \phi\sub{\rr_1}{\rr_1 \app y_1}\sub{\rr_2}{\rr_2 \app y_2}}$}
			\DP 
			\\ 
			\vvskip
			\AXC{$\Gamma \mid \Psi \p t_1 : \sigma_1 \rightarrow \tau_1 \sim t_2 : \sigma_2 \rightarrow \tau_2 \set{\forall y_1,y_2. \phi'\sub{\rr_1}{y_1}\sub{\rr_2}{y_2} \Rightarrow \phi\sub{\rr_1}{\rr_1 \app y_1}\sub{\rr_2}{\rr_2 \app y_2}} $}
			\noLine 
			\UIC{$\Gamma \mid \Psi \p u_1 : \sigma_1 \sim u_2 : \sigma_2 \set{\phi'}$}
			\UIC{$\Gamma \mid \Psi \p t_1 \app u_1 : \tau_1 \sim t_2 \app u_2 : \tau_2 \set{\phi\sub{y_1}{u_1}\sub{y_2}{u_2}}$}
			\DP
			\\ 
			\vvskip 
			\AXC{$\Gamma \mid \Psi \p t_1 : \tau_1 \sim t_2 : \tau_2 \set{\phi'}$}
			\AXC{$\Gamma \mid \Psi \p u_1 : \sigma_1 \sim u_2 : \sigma_2 \set{\phi''}$}
			\noLine
			\BIC{$\Gamma \mid \Psi \p \forall y_1,y_2,z_1,z_2. \phi'\sub{\rr_1,\rr_2}{y_1,y_2} \Rightarrow \phi''\sub{\rr_1,\rr_2}{z_1,z_2} \Rightarrow \Phi\sub{\rr_1,\rr_2}{\pair{y_1,z_1},\pair{y_2,z_2}}$}
			\UIC{$\Gamma \mid \Psi \p \pair{t_1,u_1} : \tau_1 \times \sigma_1 \sim \pair{t_2,u_2} : \tau_2 \times \sigma_2 \set{\phi}$}
			\DP
			\\ 
			\vvskip 
			\AXC{$\Gamma \mid \Psi \p t_1 : \tau_1 \times \sigma_1 \sim t_2 : \tau_2 \times \sigma_2 \set{\phi\sub{\rr_1,\rr_2}{\pi_1(\rr_1),\pi_1(\rr_2)}}$}
			\UIC{$\Gamma \mid \Psi \p \pi_1(t_1) : \tau_1 \sim \pi_1(t_2) : \tau_2 \set{\phi}$}
			\DP
			\\ 
			\vvskip
			\AXC{$\Gamma \mid \Psi \p t_1 : \tau_1 \sim t_2 : \tau_2 \set{\phi'}$}
			\AXC{$\Gamma \mid \Psi \p \phi'\sub{\rr_1,\rr_2}{t_1,t_2} \Rightarrow \phi\sub{\rr_1,\rr_2}{t_1,t_2}$}
			\BIC{$\Gamma \mid \Psi \p t_1 : \tau_1 \sim t_2 : \tau_2 \set{\phi}$}
			\DP
			\\ 
			\vvskip 
			\AXC{$\Gamma \mid \Psi \p \phi\sub{\rr_1,\rr_2}{\und{c},t_2}$}
			\AXC{$\Gamma(\Theta_2), \Theta_2 \p t_2 : \tau_2$}
			\BIC{$\Gamma \mid \Psi \p \und{c} : \B \sim t_2 : \tau_2 \set{\phi} $}
			\DP
			\qquad 
			\AXC{$\Gamma \mid \Psi \p \phi\sub{\rr_1,\rr_2}{\und{f},t_2}$}
			\AXC{$\Gamma(\Theta_2), \Theta_2 \p t_2 : \tau_2$}
			\BIC{$\Gamma \mid \Psi \p \und{f} : \seq{\B} \rightarrow \B \sim t_2 : \tau_2 \set{\phi} $}
			\DP
		\end{center}
	\end{framed}
	\caption{Some ROHOL Rules}
	\label{f:ROHOLjudgments}
\end{figure}

}
\begin{example}[Automatic Differentiation]
Let us consider the problem of formally proving 
correctness of forward mode automatic differentiation (AD) \cite{PeytonJones2019:AD,Huot2020:Correctness} 
as treated by means of open logical relations \cite{Barthe2020:OpenLogicalRelations}.\longversion{Let us now present this system on an example: correctness of forward mode automatic differentiation (AD) \cite{Barthe2020:OpenLogicalRelations,Huot2020:Correctness}.} 
In a nutshell, AD can be presented as a mapping $D$ on terms and types such that if $\Gamma \p t : \tau$ then $D(\Gamma) \p D(t) : D(\tau)$, with $D(t)$ computing the 
derivative of $t$. \shortversion{We will not detail the transformation here (a complete description can be found in \cite{Barthe2020:OpenLogicalRelations,Huot2020:Correctness},
as well as in the long version of this paper \cite{TechReport}).} \longversion{The complete description of this transformation can be found in Figure~\ref{f:ADdefinition}.}
Proving the correctness of AD amounts to showing that for any first order term 
$\Theta \p t : \type{Real}$, where $\Theta = \xx_1 : \type{Real}, \dots, \xx_n : \type{Real}$, the (semantics of the) term $D(\Theta) \p D(t) : \type{Real} \times \type{Real}$ --- with $D(\Theta) = d\xx_1 : \type{Real} \times \type{Real}, \dots, d\xx_n : \type{Real} \times \type{Real}$ --- 
is the derivative of (the semantics of) $t$. Correctness proofs have beed recently given, 
both semantically 
 \cite{Huot2020:Correctness} and syntactically \cite{Barthe2020:OpenLogicalRelations}. 
 Here, we \longshortversion{show}{sketch} how the syntactic proof by 
 Barthe et al. \cite{Barthe2020:OpenLogicalRelations} can be 
 embedded in RHOL.
\longversion{
\begin{figure}
	\begin{framed}
		\begin{center}
			We suppose given a mapping for constants and functions, $D(\und{c}) = D\und{c}$ and $D(\und{f}) = D\und{f}$.  We then define the mapping $D$ by:
			$D(\type{Real}) = \type{Real} \times \type{Real} \qquad D(\tau_1 \times \tau_2) = D(\tau_1) \times D(\tau_2) \qquad D(\tau_1 \rightarrow \tau_2) = D(\tau_1) \rightarrow D(\tau_2)$
			\\ 
			\vvskip
			$D(\cdot) = \cdot \qquad D(x : \tau, \Gamma) = dx : D(\tau), D(\Gamma)$
			\\ 
			\vvskip 
			$D(\und{c}) = D\und{c} \qquad D(\und{f}) = D\und{f} \qquad D(x) = dx \qquad D(\la x. t) = \la dx. D(t)$
			\\ 
			\vvskip
			$D(s \app t) = (D(s)) \app (D(t)) \qquad D(\pi_i(t)) = \pi_i(D(t)) \qquad D(\pair{t_1,t_2}) = \pair{D(t_1),D(t_2)}$
		\end{center}
	\end{framed}
	\caption{Automatic Differentiation}
	\label{f:ADdefinition}
\end{figure}
} 
First,  given a proof of $\Gamma, \Theta \p t : \tau$, we obtain the judgment
$$\Gamma :: \Theta, D(\Gamma) :: D(\Theta) \mid \Psi \p t : \tau \sim D(t) : D(\tau) \set{\rr_1 \LRRTh{\tau} \rr_2}$$
where $\Psi$ is a set of assertions relating elements of $\Gamma$ and $D(\Gamma)$ and $\rr_1 \LRRTh{\tau} \rr_2$ is a formula corresponding to the logical relation by Barthe et al.\cite{Barthe2020:OpenLogicalRelations}. 
The key point to stress is that in our (ROHOL) proof the flow of the derivation follows naturally from the definition of an open logical relation,
\shortversion{as elegantly showed by the case of $\lambda$-abstractions.
For a $\lambda$-abstraction, in fact, 
we have $D(\la y : \sigma. t) = \la dy : D(\sigma). D(t)$ and $D(\sigma \rightarrow \tau) = D(\sigma) \rightarrow D(\tau)$. Consequently, we have to derive the following judgment (assuming $\Gamma \p \la y. t : \sigma \rightarrow \tau$)
$$\Gamma :: \Theta, D(\Gamma) :: D(\Theta) \mid \Psi \p \la y.t : \sigma \rightarrow \tau \sim \la dy. D(t) : D(\sigma) \rightarrow D(\tau) \set{\rr_1 \LRRTh{\sigma \rightarrow \tau} \rr_2}$$
The logical relation $t_1 \LRRTh{\sigma \rightarrow \tau} t_2$ for arrow types in this case is defined by 
$$ \forall y_1 : \sigma :: \Theta. \forall y_2 : D(\sigma) :: D(\Theta). (y_1 \LRRTh{\sigma} y_2) \Rightarrow (t_1 \app y_1) \LRRTh{\tau} (t_2 \app y_2)$$
and this is exactly the shape of the $\la$-abstraction rule in ROHOL. Thus, we can apply the  rule:
\begin{center}
	\footnotesize
	\AXC{$\Gamma :: \Theta, y : \sigma :: \Theta, D(\Gamma) :: D(\Theta), dy : D(\sigma) :: D(\Theta) \mid \Psi, (y_1 \LRRTh{\sigma} y_2) \p t : \tau \sim D(t) : D(\tau) \set{\rr_1 \LRRTh{\tau} \rr_2}$}
	\UIC{$\Gamma :: \Theta, D(\Gamma) :: D(\Theta) \mid \Psi \p \la y.t : \sigma \rightarrow \tau \sim \la dy. D(t) : D(\sigma) \rightarrow D(\tau) \set{\rr_1 \LRRTh{\sigma \rightarrow \tau} \rr_2}$}
	\DP
\end{center}
and conclude the proof by induction. 
}
\longversion{ We start by defining formally the formula corresponding to the logical relation. We will use a predicate $R$ of arity $(\type{Real} :: \Theta,\type{Real} \times \type{Real} :: D(\Theta))$, that will correspond intuitively to the assessment "$R(t_1,t_2)$ is true if and only if $\sem{t_2}$ is the derivative of $\sem{t_1}$"  
\begin{definition}[Binary Open Logical Relation]
		Let $\Theta = \xx_1 : \type{Real}, \dots, \xx_n : \type{Real}$ be a fixed environment. We define the formula $t_1 \LRRTh{\tau} t_2$ by induction on the type $\tau$:
		\begin{itemize}
			\item $t_1 \LRRTh{\type{Real}} t_2 \doteq R(t_1,t_2)$
			\item $t_1 \LRRTh{\sigma \rightarrow \tau} t_2 \doteq \forall y_1 : \sigma :: \Theta, y_2 : D(\sigma) :: D(\Theta). (y_1 \LRRTh{\sigma} y_2) \Rightarrow ((t_1 \app y_1) \LRRTh{\tau} (t_2 \app y_2))$
			\item $t_1 \LRRTh{\tau_1 \times \tau_2} t_2 \doteq (\fst(t_1) \LRRTh{\tau_1} \fst(t_2)) \land (\snd(t_1) \LRRTh{\tau_2} \snd(t_2))$ 
		\end{itemize}
		And, we have that if $\Theta, \Gamma \p t_1 : \tau$ and $D(\Theta), D(\Gamma) \p t_2$ then the assertion $\Gamma :: \Theta, D(\Gamma) :: D(\Theta) \p t_1 \LRRTh{\tau} t_2$ is well-typed. 
\end{definition}
	
	Given a context $\Gamma = y_1 : \sigma_1, \dots, y_n : \sigma_n$ (and its transformation $D(\Gamma) = dy_1 : D(\sigma_1), \dots, dy_n : D(\sigma_n)$), we also introduce the following set of assertions : $\Gamma \LRRTh{\seq{\sigma}} D(\Gamma) \doteq \set{y_i \LRRTh{\sigma} dy_i \mid 1 \le i \le n}$, 
	And now, we will prove the following lemma, corresponding to the usual fundamental lemma of logical relations.
	
	\begin{lemma}[Fundamental Lemma]
		For any fixed environment $\Theta = \seq{\xx} : \seq{\B}$, for all context $\Gamma = \seq{y} : \seq{\sigma}$, for any expression $\Gamma, \Theta \p t : \tau$, then the following ROHOL judgment is valid: 
		$$\Gamma, D(\Gamma) \mid (\Gamma \LRRTh{\seq{\sigma}} D(\Gamma)) \p t : \tau \sim D(t) : D(\tau) \set{\rr_1 \LRRTh{\tau} \rr_2}$$
	\end{lemma} 
	
	Note that from this lemma, it is easy to see that first order term satisfy the correctness of AD: the second term is the derivative of the first one, by definition of the relation $R$. 
	
	\begin{proof}
		We proceed by induction on $\Gamma, \Theta \p t : \tau$. We only detail cases that do not rely on axioms for the relation $R$. Indeed, some cases are obviously inherent to $R$, for example the derivative of a variable $\xx : \type{Real}$ into $d\xx : \type{Real}$, or that the derivative of a constant is zero \dots In fact, all bases cases correspond to the use of an axiom, and it means intuitively that the AD must be valid for base types. So, we will only focus on other constructors, this way showing what we can easily derive in RHOL without using any axioms.  
		\begin{itemize}
			\item Suppose we have the typing 
			\begin{prooftree}
				\AXC{}
				\UIC{$\Gamma, \Theta \p y_i : \sigma_i$}
			\end{prooftree}
			Then, we have the following ROHOL derivation: 
			\begin{prooftree}
				\AXC{$\Gamma, D(\Gamma) \mid (\Gamma \LRRTh{\seq{\sigma}} D(\Gamma)) \p y_i \LRRTh{\sigma_i} dy_i$}
				\UIC{$\Gamma; D(\Gamma) \mid (\Gamma \LRRTh{\seq{\sigma}} D(\Gamma) \p y_i : \sigma_i \sim dy_i : D(\sigma_i) \set{\rr^1 \LRRTh{\sigma_i} \rr^2}$}
			\end{prooftree}
			And this top judgment is valid in O2HOL using a standard axiom rule as the assertion $y_i \LRRTh{\sigma_i} dy_i$ is in the set $\Gamma \LRRTh{\seq{\sigma}} d\Gamma$. 
			\item Suppose we have the typing 
			\begin{prooftree}
				\AXC{$\Gamma, y : \sigma, \Theta \p t : \tau $}
				\UIC{$\Gamma,\Theta \p \la y. t : \sigma \rightarrow \tau$}
			\end{prooftree}
			Then, we have the following ROHOL derivation: 
			\begin{prooftree}
				\footnotesize
				\AXC{Induction Hypothesis}
				\UIC{$\Gamma, D(\Gamma), y : \sigma :: \Theta, dy: D(\sigma) :: D(\Theta) \mid (\Gamma \LRRTh{\seq{\sigma}} D(\Gamma)), (y \LRRTh{\sigma} dy) \p t : \tau \sim D(t) : D(\tau) \set{\rr_1 \LRRTh{\tau} \rr_2}$}
				\UIC{$\Gamma, D(\Gamma) \mid (\Gamma \LRRTh{\seq{\sigma}} D(\Gamma)) \p \la y. t : \sigma \rightarrow \tau \sim \la dy. D(t) : D(\sigma) \rightarrow D(\tau) \set{\forall y,dy. (y \LRRTh{\sigma} dy) \Rightarrow ((\rr_1 \app y) \LRRTh{\tau} (\rr_2 \app dy))}$}
			\end{prooftree}
			And, by definition, the formula $\forall y,dy. (y \LRRTh{\sigma} dy) \Rightarrow ((\rr^1 \app y) \LRRTh{\tau} (\rr^2 \app dy))$ is the formula $\rr^1 \LRRTh{\sigma \rightarrow \tau} \rr^2$.
			\item Suppose we have the typing 
			\begin{prooftree}
				\AXC{$\Gamma,\Theta \p t_1 : \sigma \rightarrow \tau$}
				\AXC{$\Gamma,\Theta \p t_2 : \sigma$}
				\BIC{$\Gamma,\Theta \p t_1 \app t_2 : \tau$} 
			\end{prooftree}
			Then, we have the following ROHOL derivation: 
			\begin{prooftree}
				\AXC{(1)}
				\AXC{(2)}
				\BIC{$\Gamma, D(\Gamma) \mid (\Gamma \LRRTh{\seq{\sigma}} D(\Gamma)) \p t_1 \app t_2 : \tau \sim D(t_1) \app D(t_2) : D(\tau) \set{\rr_1 \LRRTh{\tau} \rr_2}$}
			\end{prooftree}
			where (1) is:
			\begin{prooftree}
				\footnotesize
				\AXC{Induction Hypothesis and Definition of $\rr^1 \LRRTh{\sigma \rightarrow \tau} \rr^2$}
				\UIC{$\Gamma, D(\Gamma) \mid (\Gamma \LRRTh{\seq{\sigma}} D(\Gamma)) \p t_1 : \sigma \rightarrow \tau \sim D(t_1) : D(\sigma) \rightarrow D(\tau) \set{\forall y,dy. (y \LRRTh{\sigma} dy) \Rightarrow ((\rr_1 \app y) \LRRTh{\tau} (\rr_2 \app dy))}$}
			\end{prooftree}
			and (2) is 
			\begin{prooftree}
				\AXC{Induction Hypothesis}
				\UIC{$\Gamma, D(\Gamma) \mid (\Gamma \LRRTh{\seq{\sigma}} D(\Gamma)) \p t_2 : \sigma \sim D(t_2) : D(\sigma) \set{\rr_1 \LRRTh{\sigma} \rr_2}$}
			\end{prooftree}
			\item Suppose we have the typing: 
			\begin{prooftree}
				\AXC{$\Gamma,\Theta \p t_1 : \tau_1$}
				\AXC{$\Gamma,\Theta \p t_2 : \tau_2$}
				\BIC{$\Gamma,\Theta \p \pair{t_1,t_2} : \tau_1 \times \tau_2$}
			\end{prooftree}
			Then, we have the following RHOL derivation: 
			\begin{prooftree}
				\AXC{(1)}
				\AXC{(2)}
				\BIC{$\Gamma, D(\Gamma) \mid (\Gamma \LRRTh{\seq{\sigma}} D(\Gamma)) \p \pair{t_1,t_2} : \tau_1 \times \tau_2 \sim \pair{D(t_1),D(t_2)} \set{\rr_1 \LRRTh{\tau_1 \times \tau_2} \rr_2}$}
			\end{prooftree}
			where (1) is 
			\begin{prooftree}
				\AXC{Induction Hypothesis}
				\UIC{$\Gamma, D(\Gamma) \mid (\Gamma \LRRTh{\seq{\sigma}} D(\Gamma)) \p t_i : \tau_i \sim D(t_i) : d\tau_i \set{\rr_1 \LRRTh{\tau_i} \rr_2}$}
			\end{prooftree}
			and (2) is 
			\begin{prooftree}
				\AXC{}
				\UIC{$\Gamma, D(\Gamma) \mid (\Gamma \LRRTh{\seq{\sigma}} D(\Gamma)) \p \forall z_1,z_2,dz_1,dz_2. (z_1 \LRRTh{\tau_1} dz_1) \land (z_2  \LRRTh{\tau_2} dz_2) \Rightarrow (\pair{z_1,z_2} \LRRTh{\tau_1 \times \tau_2} \pair{dz_1,dz_2})$}
			\end{prooftree}
			We only have to verify that this assertion is verified. We first remark that, by definition: 
			$$ \p (\pair{z_1,z_2} \LRRTh{\tau_1 \times \tau_2} \pair{dz_1,dz_2}) \Leftrightarrow (\fst(\pair{z_1,z_2}) \LRRTh{\tau_1} \fst(\pair{dz_1,dz_2})) \land (\snd(\pair{z_1,z_2}) \LRRTh{\tau_2} \snd(\pair{dz_1,dz_2}))$$
			Then, by subject conversion for HOL, we can replace $\pi_i(\pair{z_1,z_2})$ by $z_i$ and similarly for $dz_i$ and we obtain:
			$$ \p (\pair{z_1,z_2} \LRRTh{\tau_1 \times \tau_2} \pair{dz_1,dz_2}) \Leftrightarrow (z_1 \LRRTh{\tau_1} dz_1) \land (z_2 \LRRTh{\tau_2} dz_2)$$
			And from this, we can see immediately that the following formula is valid: 
			$$\p \forall z_1,z_2,dz_1,dz_2 (z_1 \LRRTh{\tau_1} dz_1) \land (z_2  \LRRTh{\tau_2} dz_2) \Rightarrow (\pair{z_1,z_2} \LRRTh{\tau_1 \times \tau_2} \pair{dz_1,dz_2})$$
			\item Suppose we have the typing (the case for $\snd$ is similar)
			\begin{prooftree}
				\AXC{$\Gamma,\Theta \p t : \tau_1 \times \tau_2$}
				\UIC{$\Gamma,\Theta \p \fst(t) : \tau_1$}
			\end{prooftree}
			Then, we can give the following RHOL derivation: 
			\begin{prooftree}
				\AXC{(1)}
				\AXC{(2)}
				\BIC{$\Gamma, D(\Gamma) \mid (\Gamma \LRRTh{\seq{\sigma}} D(\Gamma)) \p t : \tau_1 \times \tau_2 \sim D(t) : D(\tau_1) \times D(\tau_2) \set{\fst(\rr_1) \LRRTh{\tau_1} \fst(\rr_2)}$}
				\UIC{$\Gamma, D(\Gamma) \mid (\Gamma \LRRTh{\seq{\sigma}} D(\Gamma)) \p \fst(t) : \tau_1 \sim \fst(D(t)) \set{\rr_1 \LRRTh{\tau_1} \rr_2}$}
			\end{prooftree}
			where (1) is 
			\begin{prooftree}
				\AXC{Induction Hypothesis}
				\UIC{$\Gamma, D(\Gamma) \mid (\Gamma \LRRTh{\seq{\sigma}} D(\Gamma)) \p t : \tau_1 \times \tau_2 \sim D(t) : D(\tau_1) \times D(\tau_2) \set{\rr_1 \LRRTh{\tau_1 \times \tau_2} \rr_2}$}
			\end{prooftree}
			and (2) is 
			\begin{prooftree}
				\AXC{Trivial by Definition of $t \LRRTh{\tau_1 \times \tau_2} D(t)$}
				\UIC{$\Gamma, D(\Gamma) \mid (\Gamma \LRRTh{\seq{\sigma}} D(\Gamma)) \p (t \LRRTh{\tau_1 \times \tau_2} D(t)) \Rightarrow (\fst(t) \LRRTh{\tau_1} \fst(D(t)))$}
			\end{prooftree}
		\end{itemize}	
		This concludes the proof. 
	\end{proof}
}
\end{example}


\section{Local Open Higher-Order Logic}
\label{s:LOHOL}

In this section, we go beyond OHOL and introduce a \emph{local} version of OHOL
 allowing formulas to account for the domain of definition of functions. 
The motivation behind the introduction of such a logic is ultimately found in the interaction 
between the openness of OHOL --- whereby terms of a ground type are actually regarded 
as first-order functions --- and constructs of the language, whose correct semantic 
behavior relies 
on having actual terms of ground type --- and not first-order functions --- as arguments. 
To clarify, let us consider \emph{the} main construct badly interacting with the current 
formulation of OHOL: the conditional. Let us consider the term construct $\ifthen{t}{u_1}{u_0}$. The standard UHOL rule for this conditional would be:
\begin{prooftree}
	\AXC{$\Gamma \p t : \type{Bool}$}
	\AXC{$\Gamma \mid \Psi, t = \true \p u_1 : \tau \set{\phi} $}
	\AXC{$\Gamma \mid \Psi, t = \false \p u_0 : \tau \set{\phi} $}
	\TIC{$\Gamma \mid \Psi \p \ifthen{t}{u_1}{u_0} : \tau \set{\phi}$}
\end{prooftree}
expressing that this conditional satisfies a formula $\phi$ when both branches satisfy $\phi$ with the obvious assumption that in the first branch $t$ is true and in the second branch $t$ is false.  

If we naively generalize this rule to UOHOL, we see that $t =^{\Theta} \true$ does not 
describe equality between the boolean \emph{values} $t$ and $\true$: instead, it gives equality 
between the boolean \emph{function} $t$ from $\sem{\Theta}$ to $\sem{\type{Bool}}$ and the boolean \emph{constant function} $\true$ from $\sem{\Theta}$ to $\sem{\type{Bool}}$. 
Consequently, the resulting rule would be unsound, as in the first branch we cannot assume  $t$ to be constantly equal to true (there are subsets of $\Theta$ in which $t$ is true and others in which $t$ is false). 

To overcome this problem, we extend OHOL to take into account restricted  (sub)domains of $\sem{\Theta}$. We do so by defining a new logic, \emph{Local OHOL} (\LOHOL, for short).  
Compared to OHOL, the main novelty of \LOHOL{} is the introduction of formulas of 
the form $\boxx{S} \phi$, where $S$ represents a subset of $\sem{\Theta}$, 
expressing that $\phi$ is a formula on functions with domain of definition $S$, instead of the whole $\sem{\Theta}$. 
\begin{definition}
LOHOL's formulas are defined by the following grammar 
(we define disjunction, implication, and existential quantification by way of De Morgan's laws). 
\begin{align*}
	\phi &::= P^{\Theta}(S_1,\dots,S_k;t_1,\dots,t_m) \midd \bot \midd \neg \phi 
	 \midd \phi \land \phi 
	\midd \forall y: \tau. \phi \midd \boxx{S} \phi \midd \forall X. \phi
	\\
	S &::= X \midd S \cup S \midd S \cap S \midd \emptyset \midd  \complem{S} \midd \compreh{P^S(t_1,\dots,t_m)}.
\end{align*}
Here, $X$ is a set variable, $P^S(t_1,\dots,t_m)$  a predicate on terms that can use  variables in $\Theta$, and $P^{\Theta}(S_1,\dots,S_k;t_1,\dots,t_m)$ a predicate on arbitrary numbers of sets and terms. 
\end{definition} 
We denote by $(k;\tau_1,\dots,\tau_m)$ the arity of a predicate $P^{\Theta}$, with $k$ the number of sets and $\tau_i$ type of $t_i$. This generic definition allows us to have predicates on terms only --- as in OHOL --- as well as predicates on sets only --- such as the subset inclusion predicate. Finally, 
$\compreh{P^S(t_1,\dots,t_m)}$ is called 
the comprehension formula, as it gives a simple comprehension schema. 
For instance, $\boxx{\comprehfull{\xx : \type{Real}}{\xx \le 2}}$ restricts the domain 
of definition from $\mathbb{R}$ to $]{-}\infty,2]$.

Let us now move to LOHOL's semantics, beginning with the semantic of 
well-typed formulas.
First, we define the well-typed judgments $\seq{X} \mid \Gamma \pT S$ and $\seq{X} \mid \Gamma \pT \phi$, where $\seq{X}$ denotes a sequence of set variables occurring free in $S$ and and $\phi$. The most important rules are given in Figure~\ref{f:LOHOLwelltyped}. 
Such rules should be clear. In particular, the first rule, similar to an axiom rule, states that  in a judgment of 
the form $\seq{X} \mid \Gamma \pT S$ we cannot use set variables outside those in 
$\seq{X}$. Notice also that in the last rule, the second premise shows that restricting an already restricted domain amounts to intersecting the domains themselves. 
\begin{figure}
	\begin{framed}
		\begin{center}
			\AXC{}
			\UIC{$X,\seq{X} \mid \Gamma \pT X$}
			\DP
			\qquad 
			\AXC{$(\Gamma, \Theta \p t_i : \tau_i)_{1 \le i \le m} $}
			\AXC{$P^S$ has arity $(\tau_1,\dots,\tau_m)$}
			\BIC{$\seq{X} \mid \Gamma \pT \compreh{P^S(t_1,\dots,t_m)}$}
			\DP 
			\\
			\vvskip 
			\AXC{$(\Gamma, \Theta \p t_i : \tau_i)_{1 \le i \le m}$}
			\AXC{$(\seq{X} \mid \Gamma \pT S_j)_{1 \le j \le k}$}
			\AXC{$P^{\Theta}$ has arity $(k;\tau_1,\dots,\tau_m)$}
			\TIC{$\seq{X} \mid \Gamma \pT P^{\Theta}(S_1,\dots,S_k;t_1,\dots,t_m)$}
			\DP
			\\ 
			\vvskip 
			\AXC{$\seq{X} \mid \Gamma, y : \tau \pT \phi$}
			\UIC{$\seq{X} \mid \Gamma \pT \forall y : \tau. \phi$}
			\DP 
			\quad 
			\AXC{$X, \seq{X} \mid \Gamma \pT \phi$}
			\UIC{$\seq{X} \mid \Gamma \pT \forall X. \phi$}
			\DP 
			\quad
			\AXC{$\seq{X} \mid \Gamma \pT S_2$}
			\AXC{$\seq{X} \mid \Gamma \pT \phi$}
			\BIC{$\seq{X} \mid \Gamma \pT \boxx{S_2} \phi$}
			\DP
		\end{center}
	\end{framed}
	\caption{Well-Typing Judgments}
	\label{f:LOHOLwelltyped}
\end{figure}

Next, we define the denotational semantics 
$\sem{\seq{X} \mid \Gamma \pT S} : \widetilde{2^{\sem{\Theta}}} \times \sem{\Gamma^\Theta} \rightarrow 2^{\sem{\Theta}}$
 of a well-typed set $\seq{X} \mid \Gamma \pT S$, 
as mapping  subsets of $\sem{\Theta}$ (one for each set variable in $\seq{X}$) and functions 
in $ \sem{\tau}^{\sem{\Theta}}$ to subsets of $\sem{\Theta}$. 
And finally, we define the semantics $\sem{\seq{X} \mid \Gamma \pT \phi}_S \subseteq \widetilde{2^{\sem{\Theta}}} \times \sem{\Gamma^\Theta}$ of a well-typed assertion $\seq{X} \mid \Gamma \pT \phi$ parameterized by a well-typed set $\seq{X} \mid \Gamma \pT S$ as giving the possible values for set variables in $\seq{X}$ and variables in $\Gamma$ for which the formula $\phi$, on the restricted domain represented by $S$, is true. The use of this parameter $S$ is useful for the inductive definition, but in practice for a well-typed assertion $\seq{X} \mid \Gamma \pT \phi$, its semantics is defined as $\sem{\seq{X} \mid \Gamma \pT \phi}_{\complem{\emptyset}}$, taking as parameter the set representing $\sem{\Theta}$. 
 
 \begin{definition}
 \label{def:LOHOLSemantics}
 Let
 $\sem{P^S} \subseteq \sem{\tau_1} \times \cdots \times \sem{\tau_m}$
 and $\sem{P^{\Theta}} \subseteq {2^{\sem{\Theta}}}^k \times \coprod_{S \subseteq \sem{\Theta}} \prod_{1 \le i \le m} \sem{\tau_i}^S$
be interpretations of predicates $P^S$ of arity $(\tau_1, \dots, \tau_m)$ 
  and 
  $P^{\Theta}$ of arity $(k;\tau_1, \dots, \tau_m)$, respectively. Then,
  the semantics of well-typed judgments and of well-typed assertions is 
  inductively defined thus: 
			\small 
		\begin{align*}
			& \sem{X,\seq{X} \mid \Gamma \pT X}((X,\seq{X}),\seq{y}) 
			= X
			\\
			&\sem{\seq{X} \mid \Gamma \pT \compreh{P^S(t_1,\dots,t_m)}}(\seq{X},\seq{y}) 
			= \set{\seq{x} \in \sem{\Theta} \mid \sem{P^S} \owns (\sem{\Gamma, \Theta \p t_i : \tau_i}(\seq{y}(\seq{x}),\seq{x}))_{1 \le i \le m}}
			\\
			&\sem{\seq{X} \mid \Gamma \pT S_1 \cup S_2}(\seq{X},\seq{y}) 
			= \sem{\seq{X} \mid \Gamma \pT S_1}(\seq{X},\seq{y}) \cup \sem{\seq{X} \mid \Gamma \pT S_2}(\seq{X},\seq{y})
			\\
			&\sem{\seq{X} \mid \Gamma \pT P^{\Theta}(S_1,\dots,S_k;t_1,\dots,t_m)}_S 
			= 
			\\
			& \quad \set{(\seq{X},\seq{y}) \mid \sem{P^{\Theta}} \owns ((\sem{\seq{X} \mid 	\Gamma \pT S_i}(\seq{X},\seq{y}))_{1 \le i \le k}, 
		 ((\sem{\Gamma,\Theta \p t_i : \tau_i} \Comp \seq{y})_{| \sem{\seq{X} \mid \Gamma \pT S}(\seq{X},\seq{y})})_{1 \le i \le m})}
			\\
			&\sem{\seq{X} \mid \Gamma \pT \forall y : \tau. \phi}_S 
			= \set{(\seq{X},\seq{y}) \mid \forall y \in \sem{\Theta \Rightarrow \tau}, (\seq{X},(\seq{y},y)) \in \sem{\seq{X} \mid \Gamma, y : \tau  \pT \phi}_S } 
			\\
			&\sem{\seq{X} \mid \Gamma \pT \forall X. \phi}_S 
			= \set{(\seq{X},\seq{y}) \mid \forall X \in 2^{\sem{\Theta}}, ((X,\seq{X}),\seq{y}) \in \sem{X,\seq{X} \mid \Gamma \pT \phi}_S} 
			\\
			& \sem{\seq{X} \mid \Gamma \pT \boxx{S'} \phi}_S 
			= \sem{\seq{X} \mid \Gamma \pT \phi}_{S \cap S'}.
		\end{align*}
 \end{definition}

\normalsize 
 
Let us comment on some important points of Definition~\ref{def:LOHOLSemantics}. 
First, the interpretation of set predicates $P^S(t_1,\dots,t_m)$ coincides with the interpretation of HOL predicates. This way, set predicates can work on ground terms,  as 
in $t = \true$. The interpretation of predicates $P^{\Theta}$, instead, is a subset of ${2^{\sem{\Theta}}}^k \times \coprod_{S \subseteq \sem{\Theta}} \prod_{1 \le i \le m} 
 \sem{\tau_i}^S$. The first projection ${2^{\sem{\Theta}}}^k$ represents the $k$ subsets of $\sem{\Theta}$, whereas the second projection corresponds to the type of predicates in OHOL, but for all the possible subsets of $\Theta$. As an example, 
consider $\Theta = \xx : \type{Real}$ and let $\mathcal{C}^{\Theta}$ 
be a predicate of arity $(0;\type{Real})$ for local continuity. Then, 
we would interpret $\mathcal{C}^{\Theta}$ as
$$ \sem{\mathcal{C}^{\Theta}} = \{ f : A \rightarrow \mathbb{R} \mid A \subseteq \mathbb{R} \text{ and } f \text{ is locally continuous on } A\}.$$
The most interesting case of Definition~\ref{def:LOHOLSemantics} is the one for LOHOL predicates. The key point of this definition, and the main difference with OHOL, is the restriction operation $(\sem{\Gamma,\Theta \p t_i : \tau_i} \Comp \seq{y})_{| \sem{\seq{X} \mid \Gamma \pT S}(\seq{X},\seq{y})}$. The considered function $(\sem{\Gamma,\Theta \p t_i : \tau_i} \Comp \seq{y})$ is the same as the one of OHOL but, as expected for LOHOL,
we have to restrict its domain of definition to the subset represented by $S$, 
which is given by $\sem{\seq{X} \mid \Gamma \pT S}(\seq{X},\seq{y})$.\footnote{
Notice that in a set-theoretical semantics, the notion of restriction 
indeed corresponds to the usual notion of function restrictions. If one wants to give a categorical meaning to this and consider other categories than $\catname{Set}$, 
she could consider presheaves semantics for terms.}

Definition~\ref{def:LOHOLSemantics} also shows that 
the constructor $\boxx{S}$ is semantically relevant only for atomic predicates. 
Syntactically, this implies that it is always possible to commute this constructor with other 
connectives, hence pushing it just before atomic formulas. 
\begin{proposition} 
\label{prop:box-permuations}
We have the following logical equivalences: 
\begin{center}
	$\boxx{S} \top \equiv \top$ \qquad $\boxx{S} \neg \phi \equiv \neg \boxx{S} \phi$ \qquad $\boxx{S} (\phi \land \psi) \equiv \boxx{S} \phi \land \boxx{S} \psi $ \qquad $\boxx{S} \forall y : \tau. \phi \equiv \forall y : \tau. \boxx{S} \phi$ \\ \vvskip 
	$\boxx{S} \forall X. \phi \equiv \forall X. \boxx{S} \phi$ \qquad $\boxx{S} \boxx{T} \phi \equiv \boxx{S \cap T} \phi$ \qquad $\boxx{\complem{\emptyset}} \phi \equiv \phi$
\end{center}
\end{proposition}
From Proposition~\ref{prop:box-permuations} it follows that a judgment system for LOHOL needs not have a rule covering $\boxx{S}$, as it is enough to consider atomic formulas 
of the form $\boxx{S} P^{\Theta}(\dots)$. 
\begin{remark}
In the following, we will still use the notation $\boxx{S} \phi$ to improve readability. Moreover, we will 
tacitly assume to have defined an equivalence relation on sets that we can apply in formulas, 
so to identify sets (trivially) having the same semantics, such as $S \cap (T \cap U)$ and $(S \cap T) \cap U$.
\end{remark}

We now give an inference judgment system for LOHOL. The rules are close to the ones presented in Figure~\ref{f:HOLjudgments} and are given in Figure~\ref{f:LOHOLjudgment}. A judgment has the shape $\seq{X} \mid \Gamma \mid \Psi \pT \phi$, and we ask that all formulas and sets are well-typed. The main difference with OHOL is that we have to keep track of set variables. 
For the last rule, we use the following definition.
\begin{definition}[Functional Extension]
	The pointwise extension of a predicate $P^S$ with arity $(\tau_1,\dots,\tau_m)$ 
	is  the predicate $P^S_p$ of arity $(0;\tau_1,\dots,\tau_m)$ semantically 
	interpreted by:
	\begin{center}
	$ \sem{P^S_p} = \{ (f_1,\dots,f_m) \in \prod_{1 \le i \le m} \sem{\tau_i}^S \midd S \subseteq \sem{\Theta} \land \forall \seq{x} \in S. (f_1(\seq{x}),\dots,f_m(\seq{x})) \in \sem{P^S} \} $
	\end{center}
\end{definition}

\begin{example}
The pointwise extension $\le_p $ of $\le$ on real numbers corresponds to the usual pointwise comparison (on a possibly restricted domain of definition). Notice that the last rule of Figure~\ref{f:LOHOLjudgment} proves e.g. the formula $\boxx{\xx + 3 \le 2} (\xx + 3 \le_p 2)$ showing that
 restricting the domain of definition to $]{-}\infty, {-}1]$ ensures that the function $x \mapsto x + 3$ is always smaller that the constant function $x \mapsto 2$ on this domain. 
\end{example}
\begin{figure}
	\begin{framed}
		\begin{center}
			\AXC{}
			\UIC{$\seq{X} \mid \Gamma \mid \Psi, \phi \pT \phi $}
			\DP 
			\qquad 
			\AXC{$\Gamma, \Theta \p t_i : \tau$}
			\AXC{$t_1 =_{(\rightarrow)} t_2$}
			\AXC{$\seq{X} \mid \Gamma \pT S$}
			\TIC{$\seq{X} \mid \Gamma \mid \Psi \pT \boxx{S} (t_1 = t_2)$}
			\DP
			\\
			\vvskip  
			\AXC{$\seq{X} \mid \Gamma \mid \Psi \pT \boxx{S} \phi \sub{y}{t_1}$}
			\AXC{$\seq{X} \mid \Gamma \mid \Psi \pT \boxx{S} (t_1 = t_2) $}
			\BIC{$\seq{X} \mid \Gamma \mid \Psi \pT \boxx{S} \phi \sub{y}{t_2}$}
			\DP 
			\qquad 
			\AXC{$\seq{X} \mid \Gamma, y : \tau \mid \Psi \pT \phi$}
			\UIC{$\seq{X} \mid \Gamma \mid \Psi \pT \forall y : \tau. \phi $}
			\DP 
			\\
			\vvskip 
			\AXC{$\seq{X} \mid \Gamma \mid \Psi \pT \forall y : \tau. \phi $}
			\AXC{$\Gamma, \Theta \pT t : \tau$}
			\BIC{$\seq{X} \mid \Gamma \mid \Psi \pT \phi \sub{y}{t} $}
			\DP 
			\qquad 
			\AXC{$\seq{X}, X \mid \Gamma \mid \Psi \pT \phi$}
			\UIC{$\seq{X} \mid \Gamma \mid \Psi \pT \forall X. \phi $}
			\DP 
			\\
			\vspace{-0.2cm} 
			\[
			\infer{\seq{X} \mid \Gamma \mid \Psi \pT \phi \sub{X}{S}}
			{\seq{X} \mid \Gamma \mid \Psi \pT \forall X. \phi
			&
			\seq{X} \mid \Gamma \pT S}
			\quad 
			\infer{\seq{X} \mid \Gamma \mid \Psi \pT \boxx{S \cap (\compreh{P^S(t_1,\dots,t_m)})} P^S_p(t_1,\dots,t_m)}
			{P^S_p \text{ pointwise extension of }P^S}
			\]
			\end{center}
	\end{framed}
	\caption{Inference Judgment System for LOHOL}
	\label{f:LOHOLjudgment}
\end{figure}

It is easy to show that the system defined in Figure~\ref{f:LOHOLjudgment} is indeed sound. 
\begin{theorem}
	We have $\sem{\seq{X} \mid \Gamma \pT \bigwedge_{\psi \in \Psi} \psi}_{\complem{\emptyset}} \subseteq \sem{\seq{X} \mid \Gamma \pT \phi}_{\complem{\emptyset}}$ for 
	any derivable judgment $\seq{X} \mid \Gamma \mid \Psi \pT \Phi$.
\end{theorem}

Finally, we design a framework for unary predicates, called ULOHOL,  similarly to UOHOL. Remarkably, ULOHOL allows us to give a 
satisfactory rule for conditionals. 
Most of the defining rules of ULOHOL are straightforward (but notice that, 
contrary to UOHOL, we have to keep track of set variables). 
We give some of the most relevant ones in Figure~\ref{f:ULOHOLjudgments}.
Finally, we show the equivalence with the judgment system of LOHOL. 

\begin{figure}[!t]
	\begin{framed}
		\small 
		\begin{center}
			\AXC{$\seq{X} \mid \Gamma, y : \tau \mid \Psi \pT \phi\sub{\rr}{y}$}
			\UIC{$\seq{X} \mid \Gamma, y : \tau \mid \Psi \pT y : \tau \set{\phi}$}
			\DP
			\AXC{$(\xx_i : \B_i) \in \Theta$}
			\AXC{$\seq{X} \mid \Gamma \mid \Psi \pT \phi\sub{\rr}{\xx_i}$}
			\BIC{$\seq{X} \mid \Gamma \mid \Psi \pT \xx_i : \B_i \set{\phi}$}
			\DP 
			\AXC{$\seq{X} \mid \Gamma \mid \Psi \pT \phi\sub{\rr}{\und{c}}$}
			\UIC{$\seq{X} \mid \Gamma \mid \Psi \pT \und{c} : \B \set{\phi}$}
			\DP
			\\ 
			\vvskip 
			\AXC{$\seq{X} \mid \Gamma \mid \Psi \pT \phi\sub{\rr}{\und{f}}$}
			\UIC{$\seq{X} \mid \Gamma \mid \Psi \pT \und{f} : \seq{\B} \rightarrow \B \set{\phi}$}
			\DP
			\quad
			\AXC{$\seq{X} \mid \Gamma, y : \tau \mid \Psi,\psi \pT t : \sigma \set{\phi}$}
			\UIC{$\seq{X} \mid \Gamma \mid \Psi \pT \la y. t : \tau \rightarrow \sigma \set{\forall y. \psi \Rightarrow \phi\sub{\rr}{\rr \app y}}$}
			\DP
			\\ 
			\vvskip
			\AXC{$\seq{X} \mid \Gamma \mid \Psi \pT t : \tau \rightarrow \sigma \set{\forall y. \psi\sub{\rr}{y} \Rightarrow \phi\sub{\rr}{\rr \app y}}$}
			\AXC{$\seq{X} \mid \Gamma \mid \Psi \pT u : \tau \set{\psi}$}
			\BIC{$\seq{X} \mid \Gamma \mid \Psi \pT t \app u : \sigma \set{\phi\sub{y}{u}}$}
			\DP
			\\
			\vvskip
			\AXC{$\seq{X} \mid \Gamma \mid \Psi \pT t_i : \tau_i \set{\phi_i}$}
			\AXC{$\seq{X} \mid \Gamma \mid \Psi \pT \forall y,z. \phi_1\sub{\rr}{y} \land \phi_2\sub{\rr}{z} \Rightarrow \phi\sub{\rr}{\pair{y,z}}$}
			\BIC{$\seq{X} \mid \Gamma \mid \Psi \pT \pair{t_1,t_2} : \tau_1 \times \tau_2 \set{\phi}$}
			\DP
			\\ 
			\vvskip
			\AXC{$\seq{X} \mid \Gamma \mid \Psi \pT t : \tau_1 \times \tau_2 \set{\phi\sub{\rr}{\pi_i(\rr)}}$}
			\UIC{$\seq{X} \mid \Gamma \mid \Psi \pT \pi_i(t) : \tau_i \set{\phi}$}
			\DP
			\AXC{$\seq{X} \mid \Gamma \mid \Psi \pT t : \tau \set{\psi} \qquad \seq{X} \mid \Gamma \mid \Psi \pT \psi\sub{\rr}{t} \Rightarrow \phi\sub{\rr}{t}$}
			\UIC{$\seq{X} \mid \Gamma \mid \Psi \pT t : \tau \set{\phi}$}
			\DP
			\\ 
			\vvskip 
			\AXC{$\seq{X}, X \mid \Gamma \mid \Psi \pT t : \tau \set{\phi}$}
			\UIC{$\seq{X} \mid \Gamma \mid \Psi \pT t : \tau \set{\forall X. \phi}$}
			\DP 
			\quad 
			\AXC{$\seq{X} \mid \Gamma \mid \Psi \pT t : \tau \set{\forall X. \phi}$}
			\AXC{$\seq{X} \mid \Gamma \pT S$}
			\BIC{$\seq{X} \mid \Gamma \mid \Psi \pT t : \tau \set{\phi \sub{X}{S}}$}
			\DP 
			\\ 
			\vvskip 
			\AXC{$\seq{X} \mid \Gamma \mid \Psi \pT t : \type{Bool} \set{\phi_t}$}
			\AXC{$\seq{X} \mid \Gamma \mid \Psi \pT u_1 : \tau \set{\boxx{\compreh{t = \true}} \phi_1}$}
			\noLine 
			\UIC{$\seq{X} \mid \Gamma \mid \Psi \pT u_0 : \tau \set{\boxx{\compreh{t = \false}} \phi_0}$}
			\BIC{$\seq{X} \mid \Gamma \mid \Psi \pT \ifthen{t}{u_1}{u_0} : \tau \set{\phi_t \sub{r}{t} \land \boxx{\compreh{t = \true}}\phi_1 \land \boxx{\compreh{t = \false}} \phi_0}$}
			\DP 
		\end{center}
	\end{framed}
	\caption{Some Rules of ULOHOL}
	\label{f:ULOHOLjudgments}
\end{figure}

\begin{theorem}[Equivalence between ULOHOL and LOHOL]
	For every sequence of set variables $\seq{X}$, context $\Gamma$, type $\tau$, term $t$, set of assertions $\Psi$ and assertion $\phi$, 
 $\seq{X} \mid \Gamma \mid \Psi \pT t : \tau \set{\phi}$ 
 is derivable if and only if
 $\seq{X} \mid \Gamma \mid \Psi \pT \phi\sub{\rr}{t}$ is.
\end{theorem}

\begin{proof}
	The important implication is the right-to-left one.
	The proof 
	proceeds by induction on $\seq{X} \mid \Gamma \mid \Psi \pT t : \tau \set{\phi}$ following the usual proof in UHOL. For the last rule, the desired implication 
	follows from $\boxx{t =_{\type{Bool}} \false} (t =_{\Theta \rightarrow \type{Bool}} \false)$, as pointwise equality corresponds to functional equality, and so we can substitute $t$ by $\false$ in the formula $\boxx{t = \false}\phi_0$, which is semantically equivalent to replacing $\ifthen{t}{u_1}{u_0}$ by $u_0$ (and similarly for $\phi_1$). 
\end{proof}

\subsection{LOHOL at Work}

Having introduced L(U)OHOL formally, let us now see the benefits of 
its characteristic features --- namely openness, locality, and a novel rule for 
conditionals --- on a nontrivial example: local continuity of real-valued functions. 
More specifically, we take inspiration from the work by Mazza and Pagani \cite{MazzaPagani2021:ADPCF} (where it is shown that the points of error of AD on a PCF language form a set of measure zero)
and show how our system can be used to prove 
\emph{local continuity} on \emph{open sets} of $\mathbb{R}$, 
this way getting rid of those discontinuity points that may occur due to the presence of 
conditionals. 

Let us consider the context $\Theta = x : \type{Real}$ and let $t$ be 
the term $\ifthen{(x < 3)}{u_1}{u_0}$. We want to show that $t$ is continuous on a set that excludes: \emph{(i)} the possible points of discontinuity of $u_1$; 
\emph{(ii)} the possible points of discontinuity of $u_0$; 
\emph{(iii)} the point $x=3$, for which continuity is not assured. 
To do that, we need predicates and axioms for open sets, including stability of continuity by union of open sets. 
Consequently, we introduce the predicates $\open(S)$, of arity $(1;\cdot)$, and $\cont(t)$, of arity $(0;\type{Real})$. 
Our goal is then to prove a formula of the shape $(\open(S) \land \boxx{S} \cont(t))$, 
meaning that $t$ is locally continuous on the open set $S \subseteq \mathbb{R}$. 
The defining axioms of $\open$ and $\cont$ are given in Figure~\ref{f:LOHOLaxiomaticcontinuity}, 
where we assume common predicates for sets (such as $\subseteq$) as given. 

The first axiom states that restriction has no effect on $\open$, as there are simply no function to restrict, 
whereas the remaining axioms describe standard properties of open sets and continuity. Notice that we use the formula $(x \le 3 = \true)$ 
instead of $x \le 3$. 
This is because we shall construct open sets using the boolean $\LaCF$ function $<$, and thus it 
is natural to consider the predicate $(x \le 3 = \true)$. Nonetheless, it is also possible to use the formula $x \le 3$ 
as well, although we would need to add an axiom linking the function ${<} \in \mathcal{F}$ and the set predicate $<$ of LOHOL. 
\begin{figure}
	\begin{framed}
		\small 
		\begin{align*}
			\begin{aligned}
			&\vDash \forall X,Y. \boxx{X}\open(Y) \Leftrightarrow \open(Y) &&(\mathit{ORestr}) \\
			&\vDash \open(\compreh{(x < \und{c}) = \true}) &&(\mathit{ORight}) \\
			&\vDash \open(\compreh{(x > \und{c}) = \true}) &&(\mathit{OLeft}) \\ 
			&\vDash \forall X,Y. (\open(X) \land \open(Y)) \Rightarrow (\open(X \cup Y) \land \open(X \cap Y)) &&(\mathit{OStab}) \\ 
			&\vDash \forall X,Y,z. (\boxx{X}\cont(z) \land (Y \subseteq X)) \Rightarrow \boxx{Y}\cont(z)  &&(\mathit{CSubsets}) \\
			&\vDash \forall X,Y,z. (\open(X) \land \open(Y) \land \boxx{X}\cont(z) \land \boxx{Y} \cont(z)) \Rightarrow \boxx{X \cup Y}\cont(z) &&(\mathit{COpenUnion})
			\end{aligned}
		\end{align*}
	\end{framed}
\caption{Defining axioms of the predicates $\open$ and $\cont$}
\label{f:LOHOLaxiomaticcontinuity}
\end{figure}

Let us now consider the formulas $\phi_i$, with $i \in \{0,1\}$, defined by
$$
\phi_i = \boxx{\compreh{(x < 3) = \true}}(\open(S_i) \land \boxx{S_i} \cont(\rr))
$$
and the judgments 
$\cdot \mid \cdot \pT u_i : \type{Real} \set{\phi_i}$ 
which, altogether, give derivations of the form
\begin{prooftree}
	\AXC{$\cdot \mid \cdot \pT (x < 3) : \type{Bool} \set{\phi_b}$}
	\AXC{$\cdot \mid \cdot \pT u_0 : \type{Real} \set{\phi_0}$}
	\AXC{$\cdot \mid \cdot \pT u_1 : \type{Real} \set{\phi_1}$}
	\TIC{$\cdot \mid \cdot \pT \ifthen{(x < 3)}{u_1}{u_0} : \type{Real} \set{\phi_b \land \phi_1 \land \phi_0 }$}
\end{prooftree}
The target formula of this proof is not very satisfactory, as it forces us to use the implication rule. 
This, after all, is not surprising: conditionals constitute a complex case to prove continuity, and a generic rule for 
conditional cannot be expected to work directly on such a complex case. 
We overcome this problem by deriving a new rule for the conditional taking advantages of the axiomatic theory of 
$\open$ and $\cont$.

\begin{proposition}
The following rule is derivable. 
\begin{prooftree}
	\small 
	\AXC{$\Gamma \mid \Psi \pT b : \type{Bool} \set{\open(R_0) \land \open(R_1) \land (R_0 \subseteq \compreh{b = \false}) \land (R_1 \subseteq \compreh{b = \true}) }$}
	\noLine 
	\UIC{$(\forall i \in \{0,1\})\text{ } \Gamma \mid \Psi \pT t_i : \type{Real} \set{(\open(S_i) \land \boxx{S_i} \cont(\rr))}$}
	\UIC{$\Gamma \mid \Psi \pT \ifthen{b}{t_1}{t_0} : \type{Real} \set{\open((S_0 \cap R_0) \cup (S_1 \cap R_1)) \land \boxx{(S_0 \cap R_0) \cup (S_1 \cap R_1)} \cont(\rr)}$}
\end{prooftree}
\label{prop:conditional-rule-continuity}
\end{proposition}

\begin{proof}
	First, we show
	$(\open(S_0) \land \boxx{S_0} \cont(\rr)) \Rightarrow \boxx{\compreh{b = \true}}(\open(S_0) \land \boxx{S_0} \cont(\rr)),$
	for any term $b$. Recall that the constructor $\boxx{S}$ on the top level should always be understood as pushed in front of 
	atomic formulas, so that what we actually need to prove is
	$$(\open(S_0) \land \boxx{S_0} \cont(\rr)) \Rightarrow \boxx{\compreh{b = \true}}\open(S_0) \land \boxx{S_0 \cap \compreh{b = \true}} 
	\cont(\rr))$$ 
	 This follows from the axioms (Orestr) and (CSubsets), using the axiom $X \cap Y \subseteq X$ (which 
	 we assume as a defining axiom of the predicate $\subseteq$).  
	To conclude, we prove	
	$$(\phi_b \land \phi_1 \land \phi_0) \Rightarrow \open((S_0 \cap R_0) \cup (S_1 \cap R_1) \land \boxx{(S_0 \cap R_0) \cup (S_1 \cap R_1)} \cont(\rr),$$
	where $\phi_b$ is the formula $\open(R_0) \land \open(R_1) \land (R_0 \subseteq \compreh{b = \false}) \land (R_1 \subseteq \compreh{b = \true})$. This is a consequence of the axioms in Figure~\ref{f:LOHOLaxiomaticcontinuity}: openness of
	 $(S_0 \cap R_0) \cup (S_1 \cap R_1)$ comes from (Ostab), whereas continuity of $\rr$ follows from (CSubsets) and (COpenUnion). 
\end{proof}

Armed with Proposition \ref{prop:conditional-rule-continuity}, let us come back to our main example. 
Assuming an appropriate axiomatization of the relations $<, >, \le, \ge$, we define 
$R_1 = \compreh{(x < 3) = \true}$, $R_0 = \compreh{(x > 3) = \true}$, and derive the judgment 
\small 
$$\cdot \mid \cdot \pT (x < 3) : \type{Bool} \set{\open(R_0) \land \open(R_1) \land (R_0 \subseteq \compreh{(x < 3) = \false}) \land (R_1 \subseteq \compreh{(x < 3) = \true})}.$$
\normalsize
Then, using the rule for conditionals, we infer the possible discontinuity points of $t$, as well as those of $u_1$ (resp. $u_0$) smaller 
(resp. greater) than $3$, and the point $\xx=3$ itself. 


\section{Conclusion}
\label{s:Conclusion}

We have introduced several systems of \emph{open} higher-order logic, 
i.e. extensions of HOL capable of dealing with first-order properties 
natively. Even if our logics do not increase the expressive power of HOL, 
they considerably improve its effectiveness, allowing for compositional 
proofs of nontrivial program properties, such as (local) continuity and 
differentiability. We have tested our formalism on several nontrivial 
examples, obtaining compositional proofs of state-of-the-art results (such as correctness of 
automatic differentiation) in a formal framework. 

The number of extensions of OHOL defined and the heterogeneity of the examples 
analyzed suggest that OHOL (and variations thereof) 
may play an important role in the formal analysis of higher-order programming languages. 
Prompted by that observation, the authors plan to investigate applications of the logical systems to programs with effects (e.g. probabilistic programs) and to intensional program analyses (such as quantitative reasoning and program complexity).

\bibliography{biblio.bib}

\end{document}